%% file: submit01.tex
\documentclass[
twocolumn
,aps
,prb
,twoside
,showpacs
]{revtex4-1}

\usepackage[utf8x]{inputenc}
\usepackage{amsmath,amssymb,amsthm}
\usepackage{mathtools}
\usepackage{xspace}
\usepackage[usenames,dvipsnames]{xcolor}
\usepackage[colorlinks,hyperindex,breaklinks]{hyperref}
\usepackage[]{caption}
\usepackage{braket}
\usepackage{hyphenat}
\usepackage{bbold}
\usepackage{enumerate}
\usepackage{tabu,booktabs}
\usepackage{tabularx}
\usepackage[verbose]{placeins}
\usepackage[hang,flushmargin]{footmisc}
\usepackage{comment}

\usepackage[ngerman,british]{babel}

\usepackage{color}
\def\tr{\textcolor{red}}
\def\tr{}
\def\tg{\textcolor{blue}}
\def\tg{}

\DeclareMathOperator{\Tr}{Tr}

\renewcommand{\arraystretch}{1.6}
\newcommand{\dd}{\mathrm d}

\newcommand{\matr}[1]{\mathbf{\underline{#1}}}
\newcommand{\matrg}[1]{\underline{\boldsymbol{#1}}}

\newcommand{\ee}{\text e}
\newcommand{\ii}{\text i}

\renewcommand{\vec}[1]{\mathbf{#1}}
\newcommand{\spacegroup}[1]{$#1$}

\newcommand{\tableheading}[1]{\textbf{#1}}
\newcommand{\name}[1]{#1}

\newcommand{\matrixtablelinesep}{0.55mm}
\newcommand{\matrixtablearraystretch}{0.9}
\newcommand{\matrixtablearraycolsep}{0.3pt}

\newcounter{todoc}

\begin{document}

\title{ Symmetry-imposed shape of linear response tensors}

\author{M. Seemann}
\author{D. K\"odderitzsch}
\email{diemo.koedderitzsch@cup.uni-muenchen.de}
\author{S. Wimmer}
\email{sebastian.wimmer@cup.uni-muenchen.de}
\author{H. Ebert}

\affiliation{Department Chemie/Phys.\,Chemie, Ludwig-Maximilians-Universit\"at
M\"unchen, Germany}

\date{\today}


\begin{abstract}
A scheme suggested in the literature to determine the symmetry-imposed shape of
 linear response tensors is revised and extended to allow for the treatment of 
more complex situations. The extended scheme is applied to discuss the shape of
 the spin conductivity tensor for all magnetic space groups. This allows in
 particular investigating the character of longitudinal as well as transverse
 spin transport for arbitrary crystal structure and magnetic order that give 
rise e.g.\ to the spin Hall, Nernst and the spin-dependent Seebeck effects. 
In addition we draw attention to a new longitudinal spin transport phenomenon 
\tg{occurring in certain nonmagnetic solids}.

\end{abstract}


\maketitle

\section{Introduction}

The shape of a linear response tensor is of central importance as it decides 
whether a physical phenomenon may occur and what anisotropy may be
expected for a solid with given crystal symmetry and magnetic order. 
A prominent and common example for this is the anomalous Hall effect in
 ferromagnetic solids, that is connected with the non-zero anti-symmetric
contributions to the electrical conductivity tensor. Accordingly, several schemes 
were suggested in the past to predict the shape of linear response tensors
 on the basis of group-theoretical arguments \tg{(for a corresponding review 
see for example Ref.~\onlinecite{Gri93})}. Among the various schemes suggested 
that of Kleiner \cite{Kle66,Kle67,Kle69}
seems to be most convincing as it is starting from the expression for 
linear response tensors as given by Kubo's linear response formalism 
and as it uses only the behavior of the involved operators under the 
appropriate space and time transformations of the relevant magnetic space group. A further appealing 
feature of Kleiner's scheme is that it does not make use of 
Onsager's relations but allows to derive them in a most 
general way.

Kleiner's scheme was originally derived having response quantities 
in mind that are connected with the perturbation as well as the response 
represented by the components of a vector operator. A more general 
starting point is adopted in this contribution to allow the treatment 
of more complex situations. As a first simple application the tensors
 representing the charge and heat transport in response to an electric
 field and thermal gradient are considered. As a more complex transport 
quantity the corresponding spin conductivity is considered for all
 magnetic space groups. Among other things this allows the discussion of 
the transverse spin transport as occurring for the spin Hall
 \cite{DP71,KMGA04} and  spin Nernst \cite{Ma10,TGFM12} effects. 
In particular it is demonstrated that these effects may be
 discussed without use of spin-projected conductivities \cite{TGFM12,WKCE13}.\\ 

\smallskip

\section{Symmetry of Response Functions} %
Within Kubo's linear response formalism, the change of the expectation 
value of an observable $\hat B_i$ due to a time-dependent perturbation
 $\hat A_j$ can be expressed by the corresponding response function \cite{Mah00}:
%
\begin{widetext}
\begin{eqnarray}
  \tau_{\hat B_i \hat A_j}(\omega, \vec H) = \int_0^\infty \dd t \: \ee^{-\ii\omega t} \int_0^\beta \dd\lambda \Tr\left(\rho(\vec H) \hat A_j \hat B_i(t+\ii\hbar\lambda; \vec H)\right)
  \label{eq:trans_coeff_BA}
\; .
\end{eqnarray}
\end{widetext}
%
Here $ \rho(\vec H) = {\ee^{-\beta\hat H(\vec H)}}/
{\Tr(\ee^{-\beta\hat H(\vec H)})} $ is the density operator for the unperturbed system,
 the operators $\hat B_i$ and $\hat A_j$ in the Heisenberg picture are assumed to be 
 the Cartesian components of a corresponding vector operator and $\vec H$ is an external magnetic field.

Eq.\ (\ref{eq:trans_coeff_BA}) was used by Kleiner \cite{Kle66} as 
the starting point to investigate the symmetry of the tensors 
$\matrg \tau$ that describe the charge and heat transport due to
 an electric field or thermal gradient.  Kleiner's scheme, however, 
is quite general and can be easily extended to deal with more complex situations.
In the following, Kleiner's scheme will be adopted to the case when 
the observable is represented by an operator product of the form  
 $\hat B_i  \hat C_j$, again with the operators  $\hat C_i$,  
$\hat B_j$, and $\hat A_k$   being  the Cartesian components of
 a vector operator. In this case the corresponding response function is obviously given by:
\begin{widetext}
\begin{align}\begin{split}
  \tau_{(\hat B_i \hat C_j) \hat A_k}(\omega, \vec H) = \int_0^\infty \dd t \: \ee^{-\ii\omega t} \int_0^\beta \dd\lambda \Tr\left(\rho(\vec H) \hat A_k \hat B_i(t+\ii\hbar\lambda; \vec H) \hat C_j(t+\ii\hbar\lambda; \vec H)\right)
  ,
  \label{eq:kleiner_transport_coefficient_three}
\end{split}\end{align}
\end{widetext}
where by using the parenthesis in the symbol~$\tau_{(\hat B_i \hat C_j) \hat A_k}$ it is made clear that it is the observable and not the perturbation consists in a product of two operators.
The shape of the response  tensor  $\matrg \tau$ in Eq.~\eqref{eq:kleiner_transport_coefficient_three}, i.e.\ the occurrence and degeneracy of non-zero elements, has to reflect the symmetry of the investigated solid. This shape can be found by considering the impact of a symmetry operation of the corresponding space group on Eq.~\eqref{eq:kleiner_transport_coefficient_three}, as this will lead to an equation connecting the elements of  $\matrg \tau$ or possibly of a complementary tensor  $\matrg \tau'$ (see below). Collecting the restrictions imposed by all symmetry operations the shape of  $\matrg \tau$ is obtained.
In this context it is important to note that the magnetic structure of the system, if present, has to be considered. In this case, the set of symmetry operations contains unitary pure spatial~($u$), but also anti-unitary symmetry operations ($a$).

The general transformation properties of the operators $ X_i =  A_i$, $ B_i$ or $ C_i$  in Eq.~\eqref{eq:kleiner_transport_coefficient_three} under unitary  ($u$) and anti-unitary symmetry operations ($a$) can be written as:
%
\begin{eqnarray}
  \label{eq:oper_trans_u}
  u \hat X_i u^{-1} &=  &\sum_j \hat X_j D^{(\hat X)}(u)_{ji} \\
  \label{eq:oper_trans_a}
  a \hat X_i a^{-1}  &=  &\sum_j \hat X_j D^{(\hat X)}(a)_{ji}
\; ,
\end{eqnarray}
%
where $\matr D^{(\hat X)}(u)$ and $\matr D^{(\hat X)}(a)$ are the Wigner $D$-matrices corresponding to the operator $\hat X$ and operation $u$ or $a$, respectively. The group properties are reflected by the following relations:
%
\begin{eqnarray}
  \label{eq:group_properties_u}
  \matr D(uu') &=& \matr D(u) \, \matr D(u') \\
  \label{eq:group_properties_a}
  \matr D(aa') &= &\matr D(a) \, \matr D(a')^*
 \;  .
\end{eqnarray}
%
For all  unitary operations $u$ the expression under the trace in Eq.~\eqref{eq:kleiner_transport_coefficient_three} can be reformulated by cyclic permutation and by inserting the factor $u^{-1}u = 1$:
%
\begin{widetext}
\begin{eqnarray}
&\Tr\left(\ee^{-\beta\hat H(\vec H)} \hat A_k \hat B_i(t+\ii\hbar\lambda; \vec H)  \hat C_j(t+\ii\hbar\lambda; \vec H)\right)
= \Tr\left(u^{-1}u \ee^{-\beta\hat H(\vec H)} u^{-1} u \hat A_k
u^{-1} u \hat B_j(t+\ii\hbar\lambda, \vec H) u^{-1} u C_i(t+\ii\hbar\lambda; \vec H)   \right)
\nonumber \\
&= \Tr\left[ \left(u \ee^{-\beta\hat H(\vec H)} u^{-1}\right)
                  \left(u \hat A_k u^{-1} \right)
                  \left(u \hat B_i(t+\ii\hbar\lambda, \vec H) u^{-1}\right)
                  \left(u \hat C_j(t+\ii\hbar\lambda, \vec H) u^{-1}\right)
                  \right]
 \;  .
  \label{eq:trace_unitary}
\end{eqnarray}
\end{widetext}
%
The four expressions  grouped in parenthesis can now be dealt with separately. The term containing $\hat A_k$ can be rewritten using Eq.~\eqref{eq:oper_trans_u}. For the term containing $\hat B_j$ one has accordingly :
%
\begin{eqnarray}
\label{eq:Bj-transform}
  u \hat B_i(t+\ii\hbar\lambda, \vec H) u^{-1} & = & \sum_m \hat B_m(t+\ii\hbar\lambda, \vec H_u)\nonumber\\ && D^{(\hat B)}(u)_{mi}
\;   ,
\end{eqnarray}
%
with $\vec H_u$  the transformed field
%
\begin{eqnarray}
u\hat H(\vec H)u^{-1} = \hat H(\vec H_u)
\end{eqnarray}
%
connected with the operation $u$.
For the term containing $ C_j(t+\ii\hbar\lambda, \vec H)$ an analogous expression is obtained. Inserting these relations into Eq.~\eqref{eq:trace_unitary} one obtains:
%
\begin{widetext}
\begin{eqnarray}
  \Tr\left(\ee^{-\beta\hat H(\vec H)}
\hat A_k
\hat B_i(t+\ii\hbar\lambda, \vec H)
\hat C_j(t+\ii\hbar\lambda, \vec H)
\right) &= &
\sum_{lmn}
 \Tr\bigg( \ee^{-\beta\hat H(\vec H_u)}
\hat A_l
\hat B_m(t+\ii\hbar\lambda,\vec H_u)
\hat C_n(t+\ii\hbar\lambda,\vec H_u)
\nonumber \\
&& \hspace{3cm}
\,  D^{(\hat A)}(u)_{lk}
\,  D^{(\hat B)}(u)_{mi}
\,  D^{(\hat C)}(u)_{nj}\bigg)
 \;  .
  \label{eq:trace_unitary_transformed}
\end{eqnarray}
\end{widetext}
%
This equation must hold for any operators
 $\hat A_k$, $\hat B_j$ and $\hat C_i$, i.e.\ also in the special case
 $\hat  A_k = \hat B_j = \hat C_i  = \mathbb 1$, leading to:
%
\begin{eqnarray}
\Tr\left(\ee^{-\beta\hat H(\vec H)}\right) = \Tr\left(\ee^{-\beta\hat H(\vec H_u)}\right)
\;.
\end{eqnarray}
%
Inserting the two last equations into
Eq.~\eqref{eq:kleiner_transport_coefficient_three}
 for the general transport coefficients, one obtains the transformation behavior of $\matrg \tau$
under a unitary symmetry operation $u$:
%
\begin{eqnarray}
  \tau_{(\hat B_i \hat C_j) \hat A_k}(\omega, \vec H)
& =&
\sum_{lmn}
  \tau_{(\hat B_m \hat C_n) \hat A_l}(\omega, \vec H_u)
\nonumber  \\
&& \hspace{-0.25cm}
\,  D^{(\hat A)}(u)_{lk}
\,  D^{(\hat B)}(u)_{mi}
\,  D^{(\hat C)}(u)_{nj}
\,  .
  \label{eq:tau_trans_u}
\end{eqnarray}
%

A similar procedure can be applied for anti-unitary operators $a$ that contain the time reversal $T$, i.e.\ that can be decomposed as $a = v T$ with $v$ a unitary operator describing a pure spatial operation. For anti-unitary operators cyclic permutation under the trace does not hold, but one may use the relation: %
\begin{eqnarray}  \Tr(a a') = \left[ \Tr(a' a)\right]^*
 \;    .
\end{eqnarray}
%
This  expression can   be used to transform
Eq.~\eqref{eq:kleiner_transport_coefficient_three} in a similar way as done for
 Eq.~\eqref{eq:trace_unitary}
leading to:
%
\begin{widetext}
\begin{eqnarray}
&\Tr\left(\ee^{-\beta\hat H(\vec H)}
\hat A_k
\hat B_i(t+\ii\hbar\lambda, \vec H)
\hat C_j(t+\ii\hbar\lambda, \vec H)
\right) =
 \Tr\left(a^{-1}a \ee^{-\beta\hat H(\vec H)}
 a^{-1} a \hat A_k
 a^{-1} a \hat B_i(t+\ii\hbar\lambda, \vec H)
 a^{-1} a \hat C_j(t+\ii\hbar\lambda, \vec H)  \right) \nonumber\\
&= \bigg[\Tr\Big[
\left(a \ee^{-\beta\hat H(\vec H)} a^{-1}\right)
\left(a \hat A_k a^{-1} \right)
\left(a \hat B_i(t+\ii\hbar\lambda, \vec H) a^{-1}\right)
\left(a \hat C_j(t+\ii\hbar\lambda, \vec H) a^{-1}\right)
\Big]\bigg]^*
\;.
  \label{eq:kleiner_trace_anti-unitary}
\end{eqnarray}
\end{widetext}
%
Of the four expressions in parenthesis,
the second one is directly given by
Eq.~\eqref{eq:oper_trans_a},
while  the first one can be rewritten by introducing
$\vec H_a$ via the definition
%
\begin{eqnarray}
 a\hat H(\vec H) a^{-1} = \hat H(\vec H_a)
\; .
\end{eqnarray}
%
Expressing the last two terms according to
%
\begin{eqnarray}
  a \hat B_i(t + \ii\hbar\lambda, \vec H) a^{-1} & = & \sum_{m} \hat B_m(-t + \ii\hbar\lambda, \vec H) \nonumber\\ && D^{(\hat B)}(a)_{mi}
\;  ,
\end{eqnarray}
%
which follows directly from the fact that $a$
contains the time reversal operation
and
inserting these  expressions
into Eq.~\eqref{eq:kleiner_trace_anti-unitary} one arrives at:
%
\begin{widetext}
\begin{eqnarray}
  \Tr\left(
\ee^{-\beta\hat H(\vec H)}
\hat A_k
\hat B_i(t+\ii\hbar\lambda, \vec H)
\hat C_j(t+\ii\hbar\lambda, \vec H)\right)
&=& \sum_{lmn} \Tr\left[
\ee^{-\beta\hat H(\vec H_a)}
\hat A_l
\hat B_m(-t+\ii\hbar\lambda,\vec H_a)
\hat C_n(-t+\ii\hbar\lambda,\vec H_a)
\right]^*
\nonumber  \\ && \hspace{3cm}
\,  D^{(\hat A)}(u)_{lk}^*
\,  D^{(\hat B)}(u)_{mi}^*
\,  D^{(\hat C)}(u)_{nj}^*
\;  .
\end{eqnarray}
\end{widetext}
%
Using the relation \cite{9783540538332}:
%
\begin{eqnarray}
  \Tr\left(\ee^{-\beta\hat H} \hat A \hat B(\tau) \hat C(\tau) \right) = \Tr\left(\ee^{-\beta\hat H} \hat A(-\tau) \hat B \hat C \right)
\end{eqnarray}
%
one arrives  at an expression that is completely analogous to Eq.~\eqref{eq:trace_unitary_transformed}:
%
\begin{widetext}
\begin{eqnarray}
  \Tr\left(\ee^{-\beta\hat H(\vec H)}
\hat A_k
\hat B_i(t+\ii\hbar\lambda, \vec H)
\hat C_j(t+\ii\hbar\lambda, \vec H)\right)
&=& \sum_{lmn} \Tr\left(\ee^{-\beta\hat H(\vec H_a)}
\hat C_n^\dagger
\hat B_m^\dagger
\hat A_l^\dagger(t+\ii\hbar\lambda,\vec H_a)\right. \nonumber \\
&& \hspace*{2cm} \left. D^{(\hat A)}(u)_{lk}^*
\, D^{(\hat B)}(u)_{mi}^*
\, D^{(\hat C)}(u)_{nj}^*
\right) \; ,
  \label{eq:kleiner_trace_anti-unitary_transformed}
\end{eqnarray}
%
\end{widetext}
%
where
%
\begin{eqnarray}
\left( \Tr L\right)^* = \Tr\left(L^\dagger\right)
\end{eqnarray}
%
with $L$ being a linear operator was used.
Again, this equation must also hold  for the special case  $\hat A = \hat B = \hat C = \mathbb 1$, thus:
%
\begin{eqnarray}
\Tr\left(\ee^{-\beta\hat H(\vec H)}\right) = \Tr\left(\ee^{-\beta\hat H(\vec H_a)}\right)
 \;  .
\end{eqnarray}
%

Finally, inserting all these relations one obtains the transformation behavior for $\matrg \tau$ as
%
\begin{eqnarray}
  \tau_{(\hat B_i \hat C_j) \hat A_k}(\omega, \vec H) & = &
 \sum_{lmn}
\tau_{\hat A_l^\dagger
      (\hat C_n^\dagger
      \hat B_m^\dagger)
       }(\omega, \vec H_a)
\nonumber  \\ && \hspace{-0.5cm}
\, D^{(\hat A)}(a)_{lk}^*
\, D^{(\hat B)}(a)_{mi}^*
\, D^{(\hat C)}(a)_{nj}^*
 \,  ,
  \label{eq:tau_trans_a}
\end{eqnarray}
%
which is  the counter part of Eq.~\eqref{eq:tau_trans_u}, but for anti-unitary operators $a$.

It is important to note that in general the tensors
$\tau_{(\hat B_i \hat C_j) \hat A_k}$ and $\tau_{\hat A_k^\dagger (\hat B_i^\dagger \hat C_j^\dagger)}$
are different objects representing different response functions which are only interrelated by Eq.~\eqref{eq:tau_trans_a}.
Accordingly, the symbols $\matrg \tau$ and   $\matrg \tau'$ will be used below to distinguish them.
Obviously,
the two tensors  $\matrg \tau$ and  $\matrg \tau'$  coincide only if all operators and their adjoined ones are the same, i.e.\ $\hat A_i= \hat B_i$ and so on.

\medskip

Eqs.~\eqref{eq:tau_trans_u} and~\eqref{eq:tau_trans_a} relate the elements of the tensor  $\matrg \tau$ with all the elements of   $\matrg \tau$ and  $\matrg \tau'$, respectively.
As mentioned above, these relations impose for each symmetry operation restrictions on the shape of  $\matrg \tau$ that allow to decide which elements have to be zero and which are degenerate. However, to find the final shape of $\matrg \tau$ it is not necessary to derive restrictions  for all symmetry operations of the relevant  space group. Instead, it is sufficient to use only a generating set of  symmetry operations.\cite{Kle66} Finally, as was stressed by Kleiner  \cite{Kle66},  for the application of  Eqs.~\eqref{eq:tau_trans_u} and~\eqref{eq:tau_trans_a} it is not necessary to know the explicit form of the operators $\hat A_i$, $\hat B_j$ and $\hat C_k$, but only their behavior under a symmetry operation expressed by Eqs.~\eqref{eq:oper_trans_u} and~\eqref{eq:oper_trans_a}.

\section{Applications } 

\subsection{Symmetry operations and magnetic Laue groups}\label{Cryst}

For a periodic solid, the corresponding unitary symmetry operations $u$  can be represented by the Seitz symbol \cite{LK11}:
%
\begin{eqnarray}
  u = \{ R \:|\: t \}      \; ,
  \label{eq:sym_op_spatial}
\end{eqnarray}
%
where $R$ describes a (proper or improper) rotation and $t$ describes a translation.
The application of this symmetry operation on a
three dimensional vector $\vec v$ is defined as
%
\begin{eqnarray}
                    u \, \vec v = \matr R \vec v + \vec t
\; ,
\end{eqnarray}
%
where $\matr R$ is the three dimensional matrix representation of the rotation $R$ and $\vec t$ a three dimensional translation vector. For an anti-unitary symmetry operation $a$, the time reversal operation $T$ has to be considered in addition to the spatial symmetry operations. It can be included in the Seitz symbol according to:
%
\begin{eqnarray}
  \label{eq:sym_op_general}
  a = \{R \:|\: t \}\: T
 \; .
\end{eqnarray}
%
The transformation properties of a vector  $\vec v$ under $a$ depend now on its behavior under space inversion and time reversal. A vector that reverses its orientation under space inversion is called a spatial vector (or polar vector), if it stays unaltered it is called a pseudo-vector or axial vector.

Generally, the transformation of a  vector field $\vec v(\vec r)$ under an arbitrary symmetry operation $s$  is given accordingly  by:
%
\begin{eqnarray}
  s \, \vec{v}(\vec r)
  &= \pm \matr R\: \vec v(s^{-1} \vec r)
\; ,
  \label{eq:transformation_vector}
\end{eqnarray}
%
where the sign  is determined by the behavior of $\vec v(\vec r)$ under time reversal $T$
that may by part of $s$.
On the other hand, a  pseudo-vector field $\vec v(\vec r)$ transforms  as:
%
\begin{eqnarray}
  s \, \vec{v}(\vec r)
  & = \pm \det(\matr R)\: \matr R\: \vec v(s^{-1} \vec r)
\;   .
  \label{eq:transformation_pseudovector}
\end{eqnarray}
%
An example for this is the  magnetic  field $\vec H$. As  $\vec H$  changes sign under time reversal, the minus~sign in Eq.~\eqref{eq:transformation_pseudovector} applies. In particular one has
%
\begin{eqnarray}
  I \: \vec H &=& + \vec H \\
  T \: \vec H &=& - \vec H
\end{eqnarray}
%
for the application of space  inversion $I$ and time reversal $T$. \tg{In the following, we will 
use in parallel the symbols $\bar 1$ and $1'$ for $I$ and $T$, respectively.}

\medskip

Taking into account the time reversal operation, the full symmetry of a periodic solid is represented by its magnetic space group  $\cal G$ that combines all symmetry operations of the type given in Eqs.\ (\ref{eq:sym_op_spatial}) and (\ref{eq:sym_op_general}). Altogether there are 1651~magnetic space groups that fall into three categories \cite{Lit01}:
\begin{enumerate}[(a)]
  \item $\cal G$ contains the time reversal operation $T$ as an element,
  \item $\cal G$ does not contain $T$ at all, neither as a separate element nor in a combination,
  \item $\cal G$ contains $T$ only in combination with another symmetry element.
\end{enumerate}
Only nonmagnetic solids possess  one of the 230 space groups of category~a), while magnetically ordered solids belong either to  category~b) or c). Category~b) consists of 230 space groups, isomorphic to the nonmagnetic space groups, and category~c) combines  the remaining 1191~space groups.

As the crystallographic magnetic point group of a periodic solid accounts  for the translational symmetry determined by its Bravais lattice, it is sufficient to consider only the corresponding point group operations instead of the elements of its magnetic space group  when dealing with Eqs.~\eqref{eq:tau_trans_u} and~\eqref{eq:tau_trans_a}. Under certain conditions (see below) it is possible to restrict the consideration further to the corresponding magnetic Laue group of a solid, that is generated by adding the inversion operation  $I$ to the crystallographic magnetic point group. \tr{This conventional definition deviates from the older one used by Kleiner \cite{Kle66} that derives the Laue group from the corresponding crystallographic point group by removing from each improper rotation $R = P_R\:I$ its improper part $I$. For this reason} we list in 
Tables  \ref{tab:laue-groups-cat-a} --  \ref{tab:laue-groups-cat-c} all magnetic point groups of the  three categories    together with their corresponding  magnetic Laue group. \tr{The symbol in parentheses gives in addition the magnetic Laue group as used by Kleiner \cite{Kle66}.}
\newcommand{\lauetablelinesep}{1.3mm}
%
\begin{table}[h]
  \centering
  \begin{tabular}{ll}
    \toprule {magnetic point group} & {magnetic Laue group} \\
    \midrule\spacegroup{11'}, \spacegroup{\bar 1'}, \spacegroup{\bar 11'} & \spacegroup{\bar 11'} (\spacegroup{1'})\\
    \spacegroup{21'}, \spacegroup{m1'}, \spacegroup{2/m1'}, \spacegroup{2'/m}, \spacegroup{2/m'} & \spacegroup{2/m1'} (\spacegroup{21'})\\
    \spacegroup{2221'}, \spacegroup{mm21'}, \spacegroup{m'mm}, \\ \spacegroup{mmm1'}, \spacegroup{m'm'm'} & \spacegroup{mmm1'} (\spacegroup{2221'})\\
    \spacegroup{41'}, \spacegroup{\bar 41'}, \spacegroup{4/m'}, \spacegroup{4/m1'}, \spacegroup{4'/m'} & \spacegroup{4/m1'} (\spacegroup{41'})\\
    \spacegroup{4221'}, \spacegroup{4mm1'}, \spacegroup{\bar 42m1'}, \\ \spacegroup{\bar 4m21'}, \spacegroup{4/m'mm}, \spacegroup{4'/m'm'm}, \\ \spacegroup{4/mmm1'}, \spacegroup{4'/m'mm'}, \spacegroup{4/m'm'm'} & \spacegroup{4/mmm1'} (\spacegroup{42221'})\\
    \spacegroup{31'}, \spacegroup{\bar 3'}, \spacegroup{\bar 31'} & \spacegroup{\bar 31'} (\spacegroup{3'})\\
    \spacegroup{3121'}, \spacegroup{31m1'}, \spacegroup{\bar 3'1m}, \spacegroup{\bar 3'1m'}, \spacegroup{\bar 31m1'} & \spacegroup{\bar 31m1'} (\spacegroup{3'2})\\
    \spacegroup{3211'}, \spacegroup{3m11'}, \spacegroup{\bar 3'm1}, \spacegroup{\bar 3'm'1}, \spacegroup{\bar 3m11'} & \spacegroup{\bar 3m11'} (\spacegroup{3'2})\\
    \spacegroup{61'}, \spacegroup{\bar 61'}, \spacegroup{6'/m}, \spacegroup{6/m'}, \spacegroup{6/m1'} & \spacegroup{6/m1'} (\spacegroup{61'})\\
    \spacegroup{6221'}, \spacegroup{6mm1'}, \spacegroup{\bar 6m21'}, \\ \spacegroup{\bar 62m1'}, \spacegroup{6/m'mm}, \spacegroup{6'/mm'm},\\ \spacegroup{6'/mmm'}, \spacegroup{6/mmm1'}, \spacegroup{6/m'm'm'} & \spacegroup{6/mmm1'} (\spacegroup{6221'})\\
    \spacegroup{231'}, \spacegroup{m'\bar 3'}, \spacegroup{m\bar 31'} & \spacegroup{m\bar 31'} (\spacegroup{23'})\\
    \spacegroup{4321'}, \spacegroup{\bar 43m1'}, \spacegroup{m'\bar 3'm}, \spacegroup{m'\bar 3'm'}, \spacegroup{m\bar 3m1'} & \spacegroup{m\bar 3m1'} (\spacegroup{43'2})\\
    \bottomrule
  \end{tabular}
  \caption{Magnetic point groups of category~a) and their corresponding
magnetic Laue group.  \tr{In parentheses the magnetic Laue group according to its
old definition used by Kleiner \cite{Kle66} is given (see text). Because equivalent 
magnetic point group and Laue group symbols have not been removed (see text) there are 
62 and 12 instead of 53 and 11, respectively, entries.}}
  \label{tab:laue-groups-cat-a}
\end{table}
%
%
\begin{table}[h]
  \centering
  \tabulinesep=\lauetablelinesep
  \begin{tabular}{ll}
    \toprule {magnetic point group} & {magnetic Laue group} \\
    \midrule\spacegroup{1}, \spacegroup{\bar 1} & \spacegroup{\bar 1} (\spacegroup{1})\\
    \spacegroup{2}, \spacegroup{m}, \spacegroup{2/m} & \spacegroup{2/m} (\spacegroup{2})\\
    \spacegroup{222}, \spacegroup{mm2}, \spacegroup{mmm} & \spacegroup{mmm} (\spacegroup{222})\\
    \spacegroup{4}, \spacegroup{\bar 4}, \spacegroup{4/m} & \spacegroup{4/m} (\spacegroup{4})\\
    \spacegroup{422}, \spacegroup{4mm}, \spacegroup{\bar 42m}, \spacegroup{\bar 4m2}, \spacegroup{4/mmm} & \spacegroup{4/mmm} (\spacegroup{422})\\
    \spacegroup{3}, \spacegroup{\bar 3} & \spacegroup{\bar 3} (\spacegroup{3})\\
    \spacegroup{312}, \spacegroup{31m}, \spacegroup{\bar 31m} & \spacegroup{\bar 31m} (\spacegroup{32})\\
    \spacegroup{321}, \spacegroup{3m1}, \spacegroup{\bar 3m1} & \spacegroup{\bar 3m1} (\spacegroup{32})\\
    \spacegroup{6}, \spacegroup{\bar 6}, \spacegroup{6/m} & \spacegroup{6/m} (\spacegroup{6})\\
    \spacegroup{622}, \spacegroup{6mm}, \spacegroup{\bar 6m2}, \spacegroup{\bar 62m}, \spacegroup{6/mmm} & \spacegroup{6/mmm} (\spacegroup{622})\\
    \spacegroup{23}, \spacegroup{m\bar 3} & \spacegroup{m\bar 3} (\spacegroup{23})\\
    \spacegroup{432}, \spacegroup{\bar 43m}, \spacegroup{m\bar 3m} & \spacegroup{m\bar 3m} (\spacegroup{432})\\
    \bottomrule
  \end{tabular}
  \caption{Magnetic point groups of category~b) and their corresponding
magnetic Laue group. \tr{In parentheses the magnetic Laue group according to its
old definition used by Kleiner \cite{Kle66} is given (see text). Because equivalent 
magnetic point group and Laue group symbols have not been removed (see text) there are 
37 and 12 instead of 32 and 11, respectively, entries.}}
  \label{tab:laue-groups-cat-b}
\end{table}
%
%
%
\begin{table}[h]
  \centering
  \tabulinesep=\lauetablelinesep
  \begin{tabular}{ll}
    \toprule {magnetic point group} & {magnetic Laue group} \\
    \midrule\spacegroup{2'}, \spacegroup{m'}, \spacegroup{2'/m'} &  \spacegroup{2'/m'} (\spacegroup{2'})\\
    \spacegroup{2'2'2}, \spacegroup{m'm2'}, \spacegroup{m'm'2}, \spacegroup{m'm'm} &  \spacegroup{m'm'm} (\spacegroup{2'2'2})\\
    \spacegroup{4'}, \spacegroup{\bar 4'}, \spacegroup{4'/m} &  \spacegroup{4'/m} (\spacegroup{4'})\\
    \spacegroup{4'2'2}, \spacegroup{4'm'm}, \spacegroup{\bar 4'2'm},  \\ \spacegroup{\bar 4'm'2}, \spacegroup{4'/mm'm} &  \spacegroup{4'/mm'm} (\spacegroup{4'22'})\\
    \spacegroup{4'22'}, \spacegroup{4'mm'}, \spacegroup{\bar 4'2m'}, \\ \spacegroup{\bar 4'm2'}, \spacegroup{4'/mmm'} &  \spacegroup{4'/mmm'} (\spacegroup{4'22'})\\
    \spacegroup{42'2'}, \spacegroup{4m'm'}, \spacegroup{\bar 42'm'}, \\ \spacegroup{\bar 4m'2'}, \spacegroup{4/mm'm'} &  \spacegroup{4/mm'm'} (\spacegroup{42'2'})\\
    \spacegroup{312'}, \spacegroup{31m'}, \spacegroup{\bar 31m'} &  \spacegroup{\bar 31m'} (\spacegroup{32'})\\
    \spacegroup{32'1}, \spacegroup{3m'1}, \spacegroup{\bar 3m'1} &  \spacegroup{\bar 3m'1} (\spacegroup{32'})\\
    \spacegroup{6'}, \spacegroup{\bar 6'}, \spacegroup{6'/m'} &  \spacegroup{6'/m'} (\spacegroup{6'})\\
    \spacegroup{6'2'2}, \spacegroup{6'm'm}, \spacegroup{\bar 6'm'2}, \\ \spacegroup{\bar 6'2'm}, \spacegroup{6'/m'm'm} &  \spacegroup{6'/m'm'm} (\spacegroup{6'22'})\\
    \spacegroup{6'22'}, \spacegroup{6'mm'}, \spacegroup{\bar 6'm2'}, \\ \spacegroup{\bar 6'2m'}, \spacegroup{6'/m'mm'} &  \spacegroup{6'/m'mm'} (\spacegroup{6'22'})\\
    \spacegroup{62'2'}, \spacegroup{6m'm'}, \spacegroup{\bar 6m'2'}, \\ \spacegroup{\bar 62'm'}, \spacegroup{6/mm'm'} &  \spacegroup{6/mm'm'} (\spacegroup{62'2'})\\
    \spacegroup{4'32'}, \spacegroup{\bar 4'3m'}, \spacegroup{m\bar 3m'} &  \spacegroup{m\bar 3m'} (\spacegroup{4'32'})\\
    \bottomrule
  \end{tabular}
  \caption{Magnetic point groups of category~c) and their corresponding
magnetic Laue group. \tr{In parentheses the magnetic Laue group according to its
old definition used by Kleiner \cite{Kle66} is given (see text). Because equivalent 
magnetic point group and Laue group symbols have not been removed (see text) there are 
52 and 13 instead of 37 and 10, respectively, entries.}}
  \label{tab:laue-groups-cat-c}
\end{table}
%

\tr{\tg{Deriving the symbols for the magnetic point and Laue groups from those for the magnetic 
space groups\if by replacing every translation with the unity operation $1$\fi, one 
may be led in some cases to two symbols that differ only concerning the 
sequence of the second and third generators (one of these might be a dummy $1$).} As this depends
 on the chosen coordinate system and because the shape of the response tensor may 
depend on this choice, both symbols are listed although being completely equivalent.
This applies to \spacegroup{\bar31m1'} and \spacegroup{\bar3m11'} of category~a), 
\spacegroup{\bar31m} and \spacegroup{\bar3m1} of category~b) and \spacegroup{4'/mm'm}
 and \spacegroup{4'/mmm'}, \spacegroup{\bar 31m'} and \spacegroup{\bar 3m'1} as well as
\spacegroup{6'/m'm'm} and \spacegroup{6'/m'mm'} of category~c).}
\tr{Furthermore, it should be noted that for the magnetic Laue groups \spacegroup{2/m1'} 
of category~a), \spacegroup{2/m} of category~b) and \spacegroup{2'/m'} of 
category~c), the coordinate system has been chosen according to \emph{cell choice 1} of 
space group \spacegroup{2/m} as documented in the \emph{International Tables of X-ray 
Crystallography} \cite{9780792365907}.}

\subsection{Thermoelectric Coefficients} 

Within linear response theory, the induced
 electric current density $\vec j$ and the heat current density $\vec q$ are given by \cite{Kle66}
%
\begin{eqnarray}
\label{eq:linear_transport_j}
\begin{pmatrix} \vec j \\ \vec q \end{pmatrix} &= \begin{pmatrix} |e|\matr L_{11} && |e|\matr L_{12} \\ -\matr L_{21} && -\matr L_{22} \end{pmatrix} \begin{pmatrix} \nabla \mu \\ \frac{1}{T} \nabla T \end{pmatrix}
 \; ,
\end{eqnarray}
%
with $e=|e|$ the elementary charge and the electrochemical potential~$\mu$ which is related to the chemical potential~$\mu_c$ and the electric potential~$\varphi$ via
\begin{eqnarray}
  \mu = \mu_c - |e| \varphi \;.
\end{eqnarray}
As explicitly demonstrated by Kleiner  \cite{Kle66} as well
as below,
the coefficients $\matr L_{ij}$ satisfy Onsager relations of
the form
\begin{eqnarray}
\label{ONSAGER1}
  \matr L_{11}(\vec H) &=& \matr L_{11}(-\vec H)\\
\label{ONSAGER2}
  \matr L_{22}(\vec H) &=& \matr L_{22}(-\vec H)\\
\label{ONSAGER3}
  \matr L_{12}(\vec H) &=& \matr L_{21}^T(-\vec H)
    \; .
\end{eqnarray}
Identifying the operators  $\hat A_i$ and $ \hat B_i$ with one of the components of the electric current density operator $\hat j$ and the heat current density operator $\hat q$ and setting  $ \hat C_i= 1$ Eqs.~\eqref{eq:tau_trans_u} and~\eqref{eq:tau_trans_a} reduce to the expressions given by Kleiner to investigate the symmetry properties of the thermoelectric coefficients $\matr L_{ij}$. His derivation will be repeated her in a modified way as we use the conventional definition for the Laue group and as the results will be used later on. 

Expressing the   electric current density operator $\hat j = - |e| \hat v$ as a product of the electronic charge $-|e|$ and the velocity operator  $\hat v$ one can see that $\hat j$ transforms as a vector that changes sign under time reversal $T$ and space inversion $I$:
%
\begin{eqnarray}
  I \: \hat j_i &=& - \hat j_i \\
  T \: \hat j_i &=& - \hat j_i
 \;  .
\end{eqnarray}
%
The same relations apply for the heat current density operator $\hat q_i$ \cite{JM80,Kle66}.
The corresponding $3\times 3$ matrix representation for
a unitary operator $u = \{ R | t \}$ and an anti-unitary operator $a = \{ R | t \} T$
to be used in
Eqs.~\eqref{eq:tau_trans_u} and~\eqref{eq:tau_trans_a}
is:
%
\begin{eqnarray}
  \matr D^{(\hat j)}(u) &= \matr D^{(\hat q)}(u)\,  & = \; \;\,\matr D(R) \\
  \matr D^{(\hat j)}(a) &= \matr D^{(\hat q)}(a)\,  & =    - \matr D(R)
\;   .
  \label{eq:operator_transformation_currents}
\end{eqnarray}
%
Eqs.~\eqref{eq:tau_trans_u} and~\eqref{eq:tau_trans_a} (with $\hat C_i=1$)
can be brought into a more convenient form by replacing every
$\matr D$ by $\matr D = \matr R^{-1}$ and $\vec H$
 by $\vec H_{u^{-1}}$ or $\vec H_{a^{-1}}$, respectively.
Thus, Eq.~\eqref{eq:tau_trans_u}
for unitary  operators $u$ simplifies to:
%
\begin{eqnarray}
\hspace{-0.5cm}  \tau_{\hat B_i \hat A_j}\left(\omega, \vec H(R)\right) = \sum_{kl} \tau_{\hat B_k \hat A_l}(\omega, \vec H) \: D(R)_{ki}\:  D(R)_{lj}
  \label{eq:tau_trans_thermgalv_unitary_whole_rotation}
\end{eqnarray}
%
and Eq.~\eqref{eq:tau_trans_a} for  anti-unitary operators $a$ to:
%
\begin{eqnarray}
\hspace{-0.5cm}  \tau_{\hat B_i \hat A_j}\left(\omega, -\vec H(R)\right) = \sum_{kl} \tau_{\hat A_l \hat B_k}(\omega, \vec H) \: D(R)_{ki} \: D(R)_{lj}
  \label{eq:tau_trans_thermgalv_anti-unitary_whole_rotation}
\;   ,
\end{eqnarray}
%
where
%
\begin{eqnarray}
  \vec H(R)_i = \sum_i R_{ij}(P_R) H_j
\;   .
\end{eqnarray}
%
Here we used the fact that  the matrices $ D(R)_{ij}$ are real and that $\vec H$ is a pseudo-vector.
A further simplification  can be achieved by splitting $R$ in a proper rotation $P_R$ and the space inversion $I$, if it is contained in $R$. Explicitly, this means that $R = P_R$ if $R$ is a proper rotation and $R = P_R\:I$ if $R$ is an improper rotation.
 For proper rotations one has $\det(R) = +1$ while for improper rotations $\det(R) = - 1$ holds.
Because the space inversion amounts to a simple multiplication with $-\mathbb 1_3$, this splitting can be expressed by:
%
\begin{eqnarray}
  \matr D(R) = \det(R)\: \matr D(P_R)
\;   .
  \label{eq:rot_matrix_split_proper_improper}
\end{eqnarray}
%
Since the matrix $\matr D(R)$ appears twice in Eq.~\eqref{eq:tau_trans_thermgalv_unitary_whole_rotation} and~\eqref{eq:tau_trans_thermgalv_anti-unitary_whole_rotation}, the two factors  $\det(R)$ compensate each other, regardless whether $R$ is a proper or an improper rotation.
Thus, the final equation for the unitary  operators is:
%
\begin{eqnarray}
  \tau_{\hat B_i \hat A_j}(\omega, \vec H) = \sum_{kl} \tau_{\hat B_k \hat A_l}(\omega, \vec H) D(P_R)_{ki} D(P_R)_{lj}
  \label{eq:tau_trans_thermgalv_unitary}
\end{eqnarray}
%
and for anti-unitary  operators:
%
\begin{eqnarray}
\hspace{-0.5cm}  \tau_{\hat B_i \hat A_j}(\omega, -\vec H) = \sum_{kl} \tau_{\hat A_l \hat B_k}(\omega, \vec H) D(P_R)_{ki}  D(P_R)_{lj}
\;   .
  \label{eq:tau_trans_thermgalv_anti-unitary}
\end{eqnarray}
%
This splitting of $R$ enables one to consider the symmetry property of the thermogalvanic coefficients of a solid on the basis of its magnetic Laue group instead of its  magnetic point group. \tr{This applies whether the conventional definition of the Laue group (see section \ref{Cryst}) is applied or that used by Kleiner \cite{Kle66}. In the latter case the removal of the ineffective inversion $I$ happens already when constructing the Laue group. In the former case, one may add improper rotations $R = P_R \: I$ where again $I$ is ineffective and $P_R$ an element of both groups. Working only with the magnetic Laue group has the obvious} advantage that less cases have to be considered (see  Table~\ref{tab:laue-groups-cat-a} --  Table~\ref{tab:laue-groups-cat-c}) as there are only  32  magnetic Laue groups, while there are  122  different crystallographic magnetic point groups.

On the basis of Eqs.~\eqref{eq:tau_trans_thermgalv_unitary} and \eqref{eq:tau_trans_thermgalv_anti-unitary} it is now rather straightforward  to give explicit forms for the response tensors $\matr L_{ij}$ in Eq.~\eqref{eq:linear_transport_j}.
For this purpose the abbreviations $\tau_{ij} = \tau_{\hat A_i \hat B_j}$, $\tau'_{ij} = \tau_{\hat B_i \hat A_j}$ and $\sigma_{ij} = \tau_{\hat A_i \hat A_j}$ will be used, where $\hat A$ and $\hat B$ can stand for $\hat j$ or $\hat q$. Accordingly,  $\matrg \tau$ and $\matrg \tau'$ represent either $\matr L_{12}$ or $\matr L_{21}$ or the other way around, and $\matrg \sigma$ represents $\matr L_{11}$ or $\matr L_{22}$, respectively, that obviously have to have the same structure.
It is interesting to note that Eq.~\eqref{eq:tau_trans_thermgalv_anti-unitary} can lead to restrictions on the tensor elements in  addition to those imposed by Eq.~\eqref{eq:tau_trans_thermgalv_unitary}. These hold even for the  tensors of type   $\matrg \tau'$.

In the case of a magnetically ordered solid
having a magnetic space group of category~b)
the restrictions to the shape of the  thermogalvanic  tensors  result
only  from the application of Eq.~\eqref{eq:tau_trans_thermgalv_unitary}
as there are no anti-unitary operations.
As a consequence, all tensors  $\matrg \sigma$,  $\matrg \tau$  and
 $\matrg \tau'$ have the same shape.
Accordingly, only  the shape of  $\matrg \tau$
is given  in Table~\ref{tab:tau_cat_b}, that
is in full agreement with Kleiner's  Table~IV\cite{Kle66}.

For magnetic space groups belonging to  category~a) or category~c)
Eq.~\eqref{eq:tau_trans_thermgalv_anti-unitary} has to be applied
in addition to Eq.~\eqref{eq:tau_trans_thermgalv_unitary}.
In general, this leads to different symmetry restrictions for the tensors
of type
$\matrg \tau'$ and $\matrg \sigma$.
The  resulting shape of the tensors  for  category~a) is  given in
Table \ref{tab:tau_cat_a}. These results
 agree with  those given by Kleiner's Table~V\cite{Kle66},
apart from those for the Laue groups \spacegroup{\bar 31'},
\spacegroup{4/m1'} and \spacegroup{6/m1'}. \tr{Since the magnetic Laue groups in 
category~a) differ from those in b) only by the time-reversal $1'$ as an element of its own,
 the tensor shapes in Table~\ref{tab:tau_cat_a} alternatively can be deduced from those in 
Table~\ref{tab:tau_cat_b} simply by considering in addition the effect of $1'$. 
In case of $\matrg \sigma$ this can lead to additional restrictions (degeneracies 
and zero elements) since $\matrg \sigma^\prime = \matrg \sigma$. For the 
thermoelectric tensor on the other hand, this just states the usual Onsager relations 
as expressed by $\tau^\prime_{ij}(\vec{H}) = \tau_{ji}(-\vec{H})$ (see Eq.~\eqref{ONSAGER3}).}
Table \ref{tab:tau_cat_c} gives the results for  category~c) that are in full 
agreement with those given by Kleiner's Table~VI \cite{Kle66}.
%
\begin{table}[h]
  \tabulinesep=\matrixtablelinesep
  \renewcommand{\arraystretch}{\matrixtablearraystretch}
  \setlength\arraycolsep{\matrixtablearraycolsep}
  \centering
  \input{tensor_cat_a.tex}
  \caption{Tensor forms for magnetic Laue groups of category~a). }
  \label{tab:tau_cat_a}
\end{table}
%
%
\begin{table}[h]
  \tabulinesep=\matrixtablelinesep
  \renewcommand{\arraystretch}{\matrixtablearraystretch}
  \setlength\arraycolsep{\matrixtablearraycolsep}
  \centering
  \input{tensor_cat_b.tex}
  \caption{Tensor forms for magnetic Laue groups of category~b). }
  \label{tab:tau_cat_b}
\end{table}
%
%
\begin{table}[h]
  \tabulinesep=\matrixtablelinesep
  \renewcommand{\arraystretch}{\matrixtablearraystretch}
  \setlength\arraycolsep{\matrixtablearraycolsep}
  \input{tensor_cat_c.tex}
  \caption{Tensor forms for magnetic Laue groups of category~c).
 \tr{The tensor forms for the groups $4'/mm'm$ and $4'/mmm'$ are related 
to each other by a rotation of the coordinate system around the principal (z) axis 
by $\pi/4$.}}
  \label{tab:tau_cat_c}
\end{table}
%
Obviously,
the results presented in Tables
 \ref{tab:tau_cat_a}  -- \ref{tab:tau_cat_c}
fulfill the Onsager relations given by
 Eqs.\ (\ref{ONSAGER1}) to (\ref{ONSAGER3}) 
that are not postulated \emph{a priori}.

Kleiner's scheme was applied here to derive the shape of the
 tensors representing homogeneous bulk systems. However, 
it may also be applied to investigate the symmetry 
restrictions on the so-called layer-resolved conductivity tensor 
$\matrg \sigma^{IJ}$  with $ I $ and $J $ labeling 
atomic layers of a two-dimensional periodic system \cite{Huh01}.
This concept may be used for example in the context
 of electrical transport in layered GMR systems \cite{BZNM95,WDK+02}
or magneto-optical properties of surface systems
\cite{HE02a,VSUW02}.
Another extension of Kleiner's scheme is the discussion of 
non-linear effects \cite{Huh01}.

\subsection{Shape of the spin conductivity tensor } 

Spin transport as reflected
for example by the spin Hall effect is usually described
by use of the spin conductivity
 tensor $\sigma_{ij}^k$
that  gives the  current density along direction $i$
 for the spin polarization
with respect to the $k$-axis
induced by an electrical field along the $j$-axis.
Within a single-particle description of the electronic
structure the Kubo-formalism leads for  $\sigma_{ij}^k$ to
an expression analogous to  the Kubo-Bastin equation \cite{BLBN71} for the electrical conductivity
 \cite{Low10,LGK+11}:
%
\begin{eqnarray}
\label{Kubo_Bastin}
  \sigma_{ij}^k &=& \frac{\ii\hbar}{V} \int_{-\infty}^\infty \dd E\: f(E)
\nonumber \\
&&
\Tr \Big\langle {\hat J_i^k} \, \frac{\dd G^+(E)}{\dd E} \, {\hat j_j} \, \delta(E-\hat H)
\nonumber \\ && \quad \quad
 - {\hat J_i^k} \, \delta(E-\hat H) \, {\hat j_j} \, \frac{\dd G^-}{\dd E} \Big\rangle_c
\;   .
\end{eqnarray}
%
Here  $\hat H$   is the Hamiltonian of the system,
 $G^+(E)$
and   $G^-(E)$
  are the corresponding retarded and advanced Green functions and $\hat j_j$
  is the ordinary electrical current density operator.
 A straightforward definition for the spin current density operator
$\hat J_i^k=\frac{1}{2}\{\hat v_i,\sigma_k\}$
consists in \tg{the anti-commutator of} the conventional velocity operator $\hat v_i$ and 
the Pauli spin matrix $\sigma_k$.\cite{SZXN06}
As the spin conductivity is caused by spin-orbit
coupling a coherent relativistic implementation of
Eq.\ \eqref{Kubo_Bastin} seems to be more appropriate.
This implies that the electrical current density operator $ \hat j_j = -|e|c \alpha_j $ is expressed in terms of the $4\times 4$ Dirac  $\alpha$-matrices \cite{Ros61}. A corresponding expression for the spin current density operator $\hat J_i^k = \hat {\cal T}_k \hat j_i $ was suggested by Vernes et al.\ \cite{VGW07} that involves the spatial part $\hat {\cal T}_k  $ of the spin  polarization operator introduced by Bargmann and Wigner \cite{BW48}
%
\begin{eqnarray}
  \hat {\cal T}_i &= \beta \Sigma_i - \frac{1}{mc} \gamma_5 \Pi_i
  \; .
  \label{eq:spin_polarization_operator}
\end{eqnarray}
%
Here $\beta$, $\gamma_5 $, $\Sigma_i$  are again standard  $4\times 4$ Dirac-matrices,
$m$ is the electron mass and $\Pi_i$ stands for the kinetic momentum  \cite{Ros61}. In fact this approach was adopted by Lowitzer \emph{et al.}\  \cite{Low10,LGK+11} when dealing with the spin Hall effect of disordered alloys. However, as mentioned above, for an investigation of the shape of a response tensor the explicit expressions for the involved operators are not relevant but only their behavior under symmetry operations.
Both definitions of $\hat J_i^k  $ given above, consist of a combination of the velocity operator $\hat j_j $ with an operator that represents the spin polarization of an electron. In contrast to  $\hat j_i$ (see  Eq.~\eqref{eq:operator_transformation_currents}),  the latter one (e.g.\  ${\cal T}_i$) transforms as a pseudo-vector which changes sign under time reversal.
Accordingly, one has for the transformation matrices
%
\begin{eqnarray}
  \matr D^{(T_i)}(u) &=& \;\;\;\det(R) \matr D(R) \\
  \matr D^{(T_i)}(a) &=&       -\det(R) \matr D(R)
 \;  ,
\end{eqnarray}
%
corresponding to Eqs.\eqref{eq:group_properties_u} and \eqref{eq:group_properties_a}.

Identifying now $\hat A_i= \hat j_i $, $\hat B_j = \hat j_i $ and $\hat C_k= \hat {\cal T}_k$ in Eqs.~\eqref{eq:tau_trans_u} and~\eqref{eq:tau_trans_a} one finds  the  behavior of $\sigma_{ij}^k$ under unitary transformations
%
\begin{eqnarray}
  \sigma_{ij}^k &=& \sum_{lmn} \det(R) \: D(R)_{li}\:  D(R)_{mj}\: D(R)_{nk} \:\sigma_{lm}^n
\end{eqnarray}
%
and under anti-unitary transformations
%
\begin{eqnarray}
 \hspace{-0.5cm} \sigma_{ij}^k &= - \sum_{lmn} \det(R) \: D(R)_{li} \: D(R)_{mj} \: D(R)_{nk} \:\sigma_{lm}^{\prime  n }
\; ,
\end{eqnarray}
respectively.
In analogy to the treatment of thermoelectric coefficients
 presented above one may again split the rotation $R$
into its proper part $P_R$ and, if present, improper part as given
 in Eq.~\eqref{eq:rot_matrix_split_proper_improper}.
The resulting equation for unitary transformations  is then:
%
\begin{eqnarray}
\hspace{-0.5cm}  \sigma_{ij}^k &= &\sum_{lmn} D(P_R)_{li} \: D(P_R)_{mj}\:  D(P_R)_{nk} \: \det(R)^4 \:\sigma_{lm}^n\\
   &=& \sum_{lmn} D(P_R)_{li} \: D(P_R)_{mj} \: D(P_R)_{nk} \:\sigma_{lm}^n
  \label{eq:sigma_spin_polarization_transformation_unitary}
\end{eqnarray}
%
and
%
\begin{eqnarray}
  \sigma_{ij}^k &=& - \sum_{lmn} D(P_R)_{li} \: D(P_R)_{mj}\: D(P_R)_{nk} \:\sigma^{\prime n}_{lm}
  \label{eq:sigma_spin_polarization_transformation_anti-unitary}
\end{eqnarray}
%
for  anti-unitary transformations, respectively. As a consequence, as found for the thermoelectric coefficients by Kleiner \cite{Kle66} also for the spin conductivity tensor it is sufficient to  consider the magnetic Laue group of the solid.

Using Eqs.\ \eqref{eq:sigma_spin_polarization_transformation_unitary} and \eqref{eq:sigma_spin_polarization_transformation_anti-unitary} the shape of the  spin conductivity tensor  was determined with the results given \tg{in the left column of} 
Tables  \ref{tab:pol_reduced_cat_a}   --  \ref{tab:pol_reduced_cat_c} for  magnetic Laue group of  category a) -- c).

 It should be noted
 that these constitute the equivalent to the generalized Onsager relations derived by Kleiner for the $\matrg \tau'$ tensors.

Because $\hat j$ and $\hat q$ have the same transformation properties and because the tensors $\tau_{(\hat B_i \hat C_j) \hat A_k}$ and $\tau_{\hat A_l^\dagger (\hat B_m^\dagger C_n^\dagger)}$ in Eq.~\eqref{eq:tau_trans_a} are different objects in both cases, the tensor shapes for tensors describing the connection between heat currents and spin currents have exactly the same shape as those tabulated in Tab.~\ref{tab:pol_reduced_cat_a},~\ref{tab:pol_reduced_cat_b} and~\ref{tab:pol_reduced_cat_c}.

For convenience, it is possible to alter the notation of these symmetry-restricted matrices in such a way that the symmetry of the tensor is easier to recognize at first sight. \if\tg{This is achieved by removing the time-reversal operator $1'$ from every anti-unitary operation (or by omittig all anti-unitary ones?).}\fi However, this reduction leads to the loss of the specific meaning, i.e.\ the generalized Onsager relations, contained in the tensors \tg{$\matrg \sigma^{\prime k}$}. The reduced tensors are tabulated for category a), b)~and c) in \tg{the right column of} 
Tables 
\ref{tab:pol_reduced_cat_a}, \ref{tab:pol_reduced_cat_b} and \ref{tab:pol_reduced_cat_c}, respectively.

As discussed in the context of the charge and heat current in response
to an electric field the corresponding operators $\hat j_i$ and $\hat
q_i$ have the same symmetry properties.
%
%
As a consequence the tensors $\matr L_{11}$ and $\matr L_{22}$ in Eq.\ (\ref{eq:linear_transport_j})
 have the same shape given by $\matrg \sigma$ in Tables 
 \ref{tab:tau_cat_a}  -- \ref{tab:tau_cat_c}. For the same reason the tensor representing the spin current induced by an thermal gradient has the same shape as that connected with an electric field with both given by Tables 
\ref{tab:pol_reduced_cat_a} --  \ref{tab:pol_reduced_cat_c}.

\begin{widetext}

\begin{table}[h]
  \tabulinesep\matrixtablelinesep
  \renewcommand{\arraystretch}{\matrixtablearraystretch}
  \setlength\arraycolsep{\matrixtablearraycolsep}
  \centering
  \input{tensor_pol_and_reduced_cat_a.tex}
  \caption{Polarization tensor forms and reduced polarization tensor forms for magnetic Laue groups of category~a).
 \tr{The tensor forms for the groups $\overline{3}1m1'$ and $\overline{3}m11'$ are related 
to each other by a rotation of the coordinate system around the principal (z) axis by $\pi/2$.}}
  \label{tab:pol_reduced_cat_a}
\end{table}

\begin{table}[h]
  \tabulinesep=\matrixtablelinesep
  \renewcommand{\arraystretch}{\matrixtablearraystretch}
  \setlength\arraycolsep{\matrixtablearraycolsep}
  \centering
  \input{tensor_pol_and_reduced_cat_b.tex}
  \caption{Polarization tensor forms and \tr{(identical)} reduced polarization tensor forms for magnetic Laue groups of category~b).
 \tr{The tensor forms for the groups $\overline{3}1m$ and $\overline{3}m1$ are related 
to each other by a rotation of the coordinate system around the principal (z) axis 
by $\pi/2$.}}
  \label{tab:pol_reduced_cat_b}
\end{table}

\clearpage

\begin{table}[h]
  \tabulinesep=\matrixtablelinesep
  \renewcommand{\arraystretch}{\matrixtablearraystretch}
  \setlength\arraycolsep{\matrixtablearraycolsep}
  \centering
  \input{tensor_pol_and_reduced_cat_c_new.tex}
  \caption{Polarization tensor forms and reduced polarization tensor forms for 
magnetic Laue groups of category~c). \tr{Note that the reduced tensor forms for 
the groups $m'm'm$ and $4'/m$ as well as for the groups $\overline{3}1m'$, $\overline{3}m'1$, and $6'/m'$ 
are identical. Moreover those of $4'/mm'm$ and $4'/mmm'$, of $\overline{3}1m'$ and 
$\overline{3}m'1$ as well as of $6'/m'm'm$ and $6'/m'mm'$ are (pairwise) related to each other
by a rotation of the coordinate system around the principal (z) axis by $\pi/4$, $\pi/2$, and $\pi/2$, respectively.}}
  \label{tab:pol_reduced_cat_c}
\end{table}
\end{widetext}

Obviously, the occurrence of anti-symmetric off-diagonal elements in the tensor $\matrg \sigma^s$ 
($s$=x, y, z) in Table \ref{tab:pol_reduced_cat_a} implies that the transverse spin 
Hall effect is, in principle, \tg{allowed by symmetry} in any paramagnetic solid. \tr{However, one has to stress 
that in case of the magnetic Laue groups \spacegroup{\bar1}, \spacegroup{2/m}, and \spacegroup{mmm1'} the 
shape of the tensor is not purely anti-symmetric.} The 
same is true for a ferromagnetic solid according to Tables \ref{tab:pol_reduced_cat_b}  
and \ref{tab:pol_reduced_cat_c}, i.e.\ the spin Hall and Nernst effects are \tg{symmetry-allowed} 
in any magnetic solid as well \tg{(again not all cases show purely anti-symmetric elements)}. 
Considering as an example a ferromagnetic cubic solid with the magnetic Laue group 
$4/mm'm'$ (e.g.\ bcc-Fe or fcc-Ni with the magnetization along z-direction) its 
spin conductivity tensor is very different 
from the form of its nonmagnetic counterpart with 
$m\bar 3m1'$. For the nonmagnetic case only the elements 
$\sigma_{ij}^k$ with $i\ne j\ne k \ne i$ are non-zero. 
In addition, these are the same for a cyclic permutation of 
$(i,j,k)$ and change the sign for an anti-cyclic one. 
\tg{For the ferromagnetic case additional off-diagonal elements
 may appear, with the degeneracies depending on the spin
 projection component $ k $, \tr{and the tensors are no longer purely anti-symmetric.}}
 In particular one notes that there are diagonal elements that imply the occurrence of a
 longitudinal spin current induced by an electric field 
that in general will depend on whether the electric field 
is along ($\sigma_{zz}^z$ ) or perpendicular 
($\sigma_{yy}^z = \sigma_{yy}^z$) the magnetization.
 These tensor elements are obviously responsible for the 
occurrence of the spin-dependent Seebeck effect
\cite{SBAW10}. Interestingly, for a nonmagnetic solid there are several 
magnetic space groups that imply a non-vanishing diagonal 
tensor element $\sigma_{ii}^k$, i.e.\ a longitudinal current 
along the direction of the applied electric field or
thermal gradient.
This was demonstrated recently by corresponding numerical work on nonmagnetic (Au$_{1-x}$Pt$_{x}$)$_{4}$Sc showing that the longitudinal spin conductivity can be comparable in magnitude to the transverse spin Hall conductivity.\cite{WSC+15} 


\subsection{Implementation} 
\label{sec:Implementation}
The symmetry restrictions imposed on the thermogalvanic tensors by Eqs.~\eqref{eq:tau_trans_thermgalv_unitary} and \eqref{eq:tau_trans_thermgalv_anti-unitary} as well as on the spin conductivity tensor by  Eqs.~\eqref{eq:sigma_spin_polarization_transformation_unitary} and \eqref{eq:sigma_spin_polarization_transformation_anti-unitary}, respectively, were determined by means of a Python script that is based on the \textit{Computational Crystallography Toolbox}, \textit{cctbx} \cite{GSMA02}. Although this library provides support only for the nonmagnetic crystallographic operations, it is also of great value when dealing with  magnetic solids. To determine the magnetic space group of a  solid all possible  magnetic space groups are simply scanned through and checked which fits to the system under investigation. The corresponding  symmetry operations  are taken from the magnetic space group data file {\tt magnetic\_data.txt} \cite{isomag_magnetic_space_group_table,Lit01}. Once the magnetic point group has been 
found, the $u$ and $a$ operators needed for an application of Eqs.~\eqref{eq:tau_trans_thermgalv_unitary} and \eqref{eq:tau_trans_thermgalv_anti-unitary} or Eqs.~\eqref{eq:sigma_spin_polarization_transformation_unitary} and \eqref{eq:sigma_spin_polarization_transformation_anti-unitary}, respectively, are fixed.
Going through all elements of the   magnetic point group leads to  a set of connecting equations between the tensor elements which can then be solved to get the  shape of the tensor.
For these symbolic calculations the \textit{SymPy} library \cite{sympy} is used. Although in principle the generators of a  magnetic point group are sufficient to obtain all symmetry restrictions, it turned out to be  more convenient to apply all symmetry operations since the \textit{cctbx} library and the magnetic space group tables do not provide a set of generators.

Finally, it should be mentioned that the results
for the spin conductivity tensor $\matrg \sigma^z$
for the spin polarization along the $z$-axis
have been checked against
the output of the
SPRKKR program package \cite{SPR-KKR6.3}
that allows to calculate this tensor on the basis
of the relativistic  Kubo formalism \cite{EKM11}.
For all investigated magnetic  Laue groups of
 category~a) (\spacegroup{\bar 11'}, \spacegroup{mm1'},
              \spacegroup{2/m1'}, \spacegroup{4/m1'},
              \spacegroup{4/mmm1'}, \spacegroup{6/mmm1'},
              \spacegroup{m\bar3m1'}), 
          b) (\spacegroup{4/m}), and
          c) (\spacegroup{2'/m'}, \spacegroup{m'm'm}, \spacegroup{4/mm'm'}, \spacegroup{\bar 3m'1}, \spacegroup{6/mm'm'})
 the   numerical results for  $\matrg \sigma^z$ were found to be completely
in line with the analytical predictions
given  in Tables \ref{tab:pol_reduced_cat_a} --
                    \ref{tab:pol_reduced_cat_c}.


\section{Summary} 

Kleiner's scheme to determine the shape of a linear response tensor have been extended to deal with more complex situations. The scheme has been used to revise the shape of the electric charge and heat conductivity tensors for all magnetic space groups. It was demonstrated that for this only the magnetic Laue group of a solid is relevant. This also holds for the spin conductivity tensor, that is used among other to discuss the longitudinal
spin-dependent Seebeck effect as well as the transverse spin Hall and Nernst effects. \tg{Results for all magnetic space groups are presented in an easily accessible way, by giving in addition to the tensors $\matrg \sigma^{\prime k}$ containing the generalized Onsager relations also the reduced tensor forms $\matrg \sigma^{k}$. Furhermore, the axis conventions of the space groups are preserved when constructing the magnetic Laue groups and therefore, although redundant, the tensor forms are given in both coordinate systems whenever there is an ambiguity.}
Interestingly, several magnetic Laue groups for nonmagnetic solids were identified that should show a new longitudinal spin transport phenomenon.\cite{WSC+15}
\tr{Finally, it should be stressed that the scheme presented here can be applied straightforwardly to
 any other response function. Examples relevant for spintronics and related fields are the response tensors 
representing spin-orbit torque, Gilbert damping or the Edelstein effect. \cite{SKW+15}}

\begin{acknowledgments}
  This    work    was   supported    financially    by   the    Deutsche
  Forschungsgemeinschaft (DFG) via the priority program SPP 1538 and the SFB
 689.
\end{acknowledgments}


%
\end{document}

%% file: tensor_cat_a.tex
\begin{tabular}{>{\centering}m{2.3cm}cc}
\toprule \tableheading{magnetic \name{Laue} group} & \tableheading{$\matrg \tau'$} & \tableheading{$\matrg \sigma$} \\[4mm]
 \midrule
\spacegroup{\bar 11'} & \begin{math} \begin{pmatrix} \tau_{xx} && \tau_{yx} && \tau_{zx}\\\tau_{xy} && \tau_{yy} && \tau_{zy}\\\tau_{xz} && \tau_{yz} && \tau_{zz} \end{pmatrix}\end{math} & \begin{math} \begin{pmatrix} \sigma_{xx} && \sigma_{xy} && \sigma_{xz}\\\sigma_{xy} && \sigma_{yy} && \sigma_{yz}\\\sigma_{xz} && \sigma_{yz} && \sigma_{zz} \end{pmatrix}\end{math}\\[4mm]
 \spacegroup{2/m1'} & \begin{math} \begin{pmatrix} \tau_{xx} && 0 && \tau_{zx}\\0 && \tau_{yy} && 0\\\tau_{xz} && 0 && \tau_{zz} \end{pmatrix}\end{math} & \begin{math} \begin{pmatrix} \sigma_{xx} && 0 && \sigma_{xz}\\0 && \sigma_{yy} && 0\\\sigma_{xz} && 0 && \sigma_{zz} \end{pmatrix}\end{math}\\[4mm]
 \spacegroup{mmm1'} & \begin{math} \begin{pmatrix} \tau_{xx} && 0 && 0\\0 && \tau_{yy} && 0\\0 && 0 && \tau_{zz} \end{pmatrix}\end{math} & \begin{math} \begin{pmatrix} \sigma_{xx} && 0 && 0\\0 && \sigma_{yy} && 0\\0 && 0 && \sigma_{zz} \end{pmatrix}\end{math}\\[4mm]
 \spacegroup{\bar 31'}, \spacegroup{4/m1'}, \spacegroup{6/m1'} & \begin{math} \begin{pmatrix} \tau_{xx} && -\tau_{xy} && 0\\\tau_{xy} && \tau_{xx} && 0\\0 && 0 && \tau_{zz} \end{pmatrix}\end{math} & \begin{math} \begin{pmatrix} \sigma_{xx} && 0 && 0\\0 && \sigma_{xx} && 0\\0 && 0 && \sigma_{zz} \end{pmatrix}\end{math}\\[4mm]
 \spacegroup{\bar 31m1'}, \spacegroup{\bar 3m11'}, \spacegroup{4/mmm1'}, \spacegroup{6/mmm1'} & \begin{math} \begin{pmatrix} \tau_{xx} && 0 && 0\\0 && \tau_{xx} && 0\\0 && 0 && \tau_{zz} \end{pmatrix}\end{math} & \begin{math} \begin{pmatrix} \sigma_{xx} && 0 && 0\\0 && \sigma_{xx} && 0\\0 && 0 && \sigma_{zz} \end{pmatrix}\end{math}\\[4mm]
 \spacegroup{m\bar 31'}, \spacegroup{m\bar 3m1'} & \begin{math} \begin{pmatrix} \tau_{xx} && 0 && 0\\0 && \tau_{xx} && 0\\0 && 0 && \tau_{xx} \end{pmatrix}\end{math} & \begin{math} \begin{pmatrix} \sigma_{xx} && 0 && 0\\0 && \sigma_{xx} && 0\\0 && 0 && \sigma_{xx} \end{pmatrix}\end{math}\\[4mm]
 \bottomrule
\end{tabular}

%% file: tensor_cat_b.tex
\begin{tabular}{lc}
\toprule \tableheading{magnetic \name{Laue} group} & \tableheading{$\matrg \tau$} \\[4mm]
 \midrule
\spacegroup{\bar 1} & \begin{math} \begin{pmatrix} \tau_{xx} && \tau_{xy} && \tau_{xz}\\\tau_{yx} && \tau_{yy} && \tau_{yz}\\\tau_{zx} && \tau_{zy} && \tau_{zz} \end{pmatrix}\end{math}\\[4mm]
 \spacegroup{2/m} & \begin{math} \begin{pmatrix} \tau_{xx} && 0 && \tau_{xz}\\0 && \tau_{yy} && 0\\\tau_{zx} && 0 && \tau_{zz} \end{pmatrix}\end{math}\\[4mm]
 \spacegroup{mmm} & \begin{math} \begin{pmatrix} \tau_{xx} && 0 && 0\\0 && \tau_{yy} && 0\\0 && 0 && \tau_{zz} \end{pmatrix}\end{math}\\[4mm]
 \spacegroup{\bar 3}, \spacegroup{4/m}, \spacegroup{6/m} & \begin{math} \begin{pmatrix} \tau_{xx} && \tau_{xy} && 0\\-\tau_{xy} && \tau_{xx} && 0\\0 && 0 && \tau_{zz} \end{pmatrix}\end{math}\\[4mm]
 \spacegroup{\bar 31m}, \spacegroup{\bar 3m1}, \spacegroup{4/mmm}, \spacegroup{6/mmm} & \begin{math} \begin{pmatrix} \tau_{xx} && 0 && 0\\0 && \tau_{xx} && 0\\0 && 0 && \tau_{zz} \end{pmatrix}\end{math}\\[4mm]
 \spacegroup{m\bar 3}, \spacegroup{m\bar 3m} & \begin{math} \begin{pmatrix} \tau_{xx} && 0 && 0\\0 && \tau_{xx} && 0\\0 && 0 && \tau_{xx} \end{pmatrix}\end{math}\\[4mm]
 \bottomrule
\end{tabular}

%% file: tensor_cat_c.tex
\begin{tabular}{>{\centering}m{2.3cm}cc}
\toprule \tableheading{magnetic \name{Laue} group} & \tableheading{$\matrg \tau'$} & \tableheading{$\matrg \sigma$} \\[4mm]
 \midrule
\spacegroup{2'/m'} & \begin{math} \begin{pmatrix} \tau_{xx} && -\tau_{yx} && \tau_{zx}\\-\tau_{xy} && \tau_{yy} && -\tau_{zy}\\\tau_{xz} && -\tau_{yz} && \tau_{zz} \end{pmatrix}\end{math} & \begin{math} \begin{pmatrix} \sigma_{xx} && \sigma_{xy} && \sigma_{xz}\\-\sigma_{xy} && \sigma_{yy} && \sigma_{yz}\\\sigma_{xz} && -\sigma_{yz} && \sigma_{zz} \end{pmatrix}\end{math}\\[4mm]
 \spacegroup{m'm'm} & \begin{math} \begin{pmatrix} \tau_{xx} && -\tau_{yx} && 0\\-\tau_{xy} && \tau_{yy} && 0\\0 && 0 && \tau_{zz} \end{pmatrix}\end{math} & \begin{math} \begin{pmatrix} \sigma_{xx} && \sigma_{xy} && 0\\-\sigma_{xy} && \sigma_{yy} && 0\\0 && 0 && \sigma_{zz} \end{pmatrix}\end{math}\\[4mm]
 \spacegroup{4'/m} & \begin{math} \begin{pmatrix} \tau_{yy} && -\tau_{xy} && 0\\-\tau_{yx} && \tau_{xx} && 0\\0 && 0 && \tau_{zz} \end{pmatrix}\end{math} & \begin{math} \begin{pmatrix} \sigma_{xx} && 0 && 0\\0 && \sigma_{xx} && 0\\0 && 0 && \sigma_{zz} \end{pmatrix}\end{math}\\[4mm]
 \spacegroup{4'/mm'm} & \begin{math} \begin{pmatrix} \tau_{xx} && -\tau_{xy} && 0\\-\tau_{xy} && \tau_{xx} && 0\\0 && 0 && \tau_{zz} \end{pmatrix}\end{math} & \begin{math} \begin{pmatrix} \sigma_{xx} && 0 && 0\\0 && \sigma_{xx} && 0\\0 && 0 && \sigma_{zz} \end{pmatrix}\end{math}\\[4mm]
 \spacegroup{4'/mmm'} & \begin{math} \begin{pmatrix} \tau_{yy} && 0 && 0\\0 && \tau_{xx} && 0\\0 && 0 && \tau_{zz} \end{pmatrix}\end{math} & \begin{math} \begin{pmatrix} \sigma_{xx} && 0 && 0\\0 && \sigma_{xx} && 0\\0 && 0 && \sigma_{zz} \end{pmatrix}\end{math}\\[4mm]
 \spacegroup{\bar 31m'}, \spacegroup{\bar 3m'1}, \spacegroup{4/mm'm'}, \spacegroup{6/mm'm'} & \begin{math} \begin{pmatrix} \tau_{xx} && \tau_{xy} && 0\\-\tau_{xy} && \tau_{xx} && 0\\0 && 0 && \tau_{zz} \end{pmatrix}\end{math} & \begin{math} \begin{pmatrix} \sigma_{xx} && \sigma_{xy} && 0\\-\sigma_{xy} && \sigma_{xx} && 0\\0 && 0 && \sigma_{zz} \end{pmatrix}\end{math}\\[4mm]
 \spacegroup{6'/m'} & \begin{math} \begin{pmatrix} \tau_{xx} && -\tau_{xy} && 0\\\tau_{xy} && \tau_{xx} && 0\\0 && 0 && \tau_{zz} \end{pmatrix}\end{math} & \begin{math} \begin{pmatrix} \sigma_{xx} && 0 && 0\\0 && \sigma_{xx} && 0\\0 && 0 && \sigma_{zz} \end{pmatrix}\end{math}\\[4mm]
 \spacegroup{6'/m'm'm}, \spacegroup{6'/m'mm'} & \begin{math} \begin{pmatrix} \tau_{xx} && 0 && 0\\0 && \tau_{xx} && 0\\0 && 0 && \tau_{zz} \end{pmatrix}\end{math} & \begin{math} \begin{pmatrix} \sigma_{xx} && 0 && 0\\0 && \sigma_{xx} && 0\\0 && 0 && \sigma_{zz} \end{pmatrix}\end{math}\\[4mm]
 \spacegroup{m\bar 3m'} & \begin{math} \begin{pmatrix} \tau_{xx} && 0 && 0\\0 && \tau_{xx} && 0\\0 && 0 && \tau_{xx} \end{pmatrix}\end{math} & \begin{math} \begin{pmatrix} \sigma_{xx} && 0 && 0\\0 && \sigma_{xx} && 0\\0 && 0 && \sigma_{xx} \end{pmatrix}\end{math}\\[4mm]
 \bottomrule
\end{tabular}

%% file: tensor_pol_and_reduced_cat_a.tex
\begin{tabular}{>{\centering}m{1.5cm}| ccc|ccc}
\toprule \tableheading{magnetic \name{Laue} group} & \tableheading{$\matrg \sigma'^x$} & \tableheading{$\matrg \sigma'^y$} & \tableheading{$\matrg \sigma'^z$} & \tableheading{$\matrg \sigma^x$} & \tableheading{$\matrg \sigma^y$} & \tableheading{$\matrg \sigma^z$} \\[4mm]
 \midrule
\spacegroup{\bar 11'} & \begin{math} \begin{pmatrix} -\sigma^{x}_{xx} && -\sigma^{x}_{xy} && -\sigma^{x}_{xz}\\-\sigma^{y}_{xx} && -\sigma^{y}_{xy} && -\sigma^{y}_{xz}\\-\sigma^{z}_{xx} && -\sigma^{z}_{xy} && -\sigma^{z}_{xz} \end{pmatrix}\end{math} & \begin{math} \begin{pmatrix} -\sigma^{x}_{yx} && -\sigma^{x}_{yy} && -\sigma^{x}_{yz}\\-\sigma^{y}_{yx} && -\sigma^{y}_{yy} && -\sigma^{y}_{yz}\\-\sigma^{z}_{yx} && -\sigma^{z}_{yy} && -\sigma^{z}_{yz} \end{pmatrix}\end{math} & \begin{math} \begin{pmatrix} -\sigma^{x}_{zx} && -\sigma^{x}_{zy} && -\sigma^{x}_{zz}\\-\sigma^{y}_{zx} && -\sigma^{y}_{zy} && -\sigma^{y}_{zz}\\-\sigma^{z}_{zx} && -\sigma^{z}_{zy} && -\sigma^{z}_{zz} \end{pmatrix}\end{math} & \begin{math} \begin{pmatrix} \sigma^{x}_{xx} && \sigma^{x}_{xy} && \sigma^{x}_{xz}\\\sigma^{x}_{yx} && \sigma^{x}_{yy} && \sigma^{x}_{yz}\\\sigma^{x}_{zx} && \sigma^{x}_{zy} && \sigma^{x}_{zz} \end{pmatrix}\end{math} & \begin{math} \begin{pmatrix} \sigma^{y}_{xx} && \sigma^{y}_{xy} && \sigma^{y}_{xz}\\\sigma^{y}_{yx} && \sigma^{y}_{yy} && \sigma^{y}_{yz}\\\sigma^{y}_{zx} && \sigma^{y}_{zy} && \sigma^{y}_{zz} \end{pmatrix}\end{math} & \begin{math} \begin{pmatrix} \sigma^{z}_{xx} && \sigma^{z}_{xy} && \sigma^{z}_{xz}\\\sigma^{z}_{yx} && \sigma^{z}_{yy} && \sigma^{z}_{yz}\\\sigma^{z}_{zx} && \sigma^{z}_{zy} && \sigma^{z}_{zz} \end{pmatrix}\end{math}\\[4mm]
 \spacegroup{2/m1'} & \begin{math} \begin{pmatrix} 0 && -\sigma^{x}_{xy} && 0\\-\sigma^{y}_{xx} && 0 && -\sigma^{y}_{xz}\\0 && -\sigma^{z}_{xy} && 0 \end{pmatrix}\end{math} & \begin{math} \begin{pmatrix} -\sigma^{x}_{yx} && 0 && -\sigma^{x}_{yz}\\0 && -\sigma^{y}_{yy} && 0\\-\sigma^{z}_{yx} && 0 && -\sigma^{z}_{yz} \end{pmatrix}\end{math} & \begin{math} \begin{pmatrix} 0 && -\sigma^{x}_{zy} && 0\\-\sigma^{y}_{zx} && 0 && -\sigma^{y}_{zz}\\0 && -\sigma^{z}_{zy} && 0 \end{pmatrix}\end{math} & \begin{math} \begin{pmatrix} 0 && \sigma^{x}_{xy} && 0\\\sigma^{x}_{yx} && 0 && \sigma^{x}_{yz}\\0 && \sigma^{x}_{zy} && 0 \end{pmatrix}\end{math} & \begin{math} \begin{pmatrix} \sigma^{y}_{xx} && 0 && \sigma^{y}_{xz}\\0 && \sigma^{y}_{yy} && 0\\\sigma^{y}_{zx} && 0 && \sigma^{y}_{zz} \end{pmatrix}\end{math} & \begin{math} \begin{pmatrix} 0 && \sigma^{z}_{xy} && 0\\\sigma^{z}_{yx} && 0 && \sigma^{z}_{yz}\\0 && \sigma^{z}_{zy} && 0 \end{pmatrix}\end{math}\\[4mm]
 \spacegroup{mmm1'} & \begin{math} \begin{pmatrix} 0 && 0 && 0\\0 && 0 && -\sigma^{y}_{xz}\\0 && -\sigma^{z}_{xy} && 0 \end{pmatrix}\end{math} & \begin{math} \begin{pmatrix} 0 && 0 && -\sigma^{x}_{yz}\\0 && 0 && 0\\-\sigma^{z}_{yx} && 0 && 0 \end{pmatrix}\end{math} & \begin{math} \begin{pmatrix} 0 && -\sigma^{x}_{zy} && 0\\-\sigma^{y}_{zx} && 0 && 0\\0 && 0 && 0 \end{pmatrix}\end{math} & \begin{math} \begin{pmatrix} 0 && 0 && 0\\0 && 0 && \sigma^{x}_{yz}\\0 && \sigma^{x}_{zy} && 0 \end{pmatrix}\end{math} & \begin{math} \begin{pmatrix} 0 && 0 && \sigma^{y}_{xz}\\0 && 0 && 0\\\sigma^{y}_{zx} && 0 && 0 \end{pmatrix}\end{math} & \begin{math} \begin{pmatrix} 0 && \sigma^{z}_{xy} && 0\\\sigma^{z}_{yx} && 0 && 0\\0 && 0 && 0 \end{pmatrix}\end{math}\\[4mm]
 \spacegroup{4/m1'}, \spacegroup{6/m1'} & \begin{math} \begin{pmatrix} 0 && 0 && -\sigma^{x}_{xz}\\0 && 0 && -\sigma^{y}_{xz}\\-\sigma^{z}_{xx} && -\sigma^{z}_{xy} && 0 \end{pmatrix}\end{math} & \begin{math} \begin{pmatrix} 0 && 0 && \sigma^{y}_{xz}\\0 && 0 && -\sigma^{x}_{xz}\\\sigma^{z}_{xy} && -\sigma^{z}_{xx} && 0 \end{pmatrix}\end{math} & \begin{math} \begin{pmatrix} -\sigma^{x}_{zx} && \sigma^{y}_{zx} && 0\\-\sigma^{y}_{zx} && -\sigma^{x}_{zx} && 0\\0 && 0 && -\sigma^{z}_{zz} \end{pmatrix}\end{math} & \begin{math} \begin{pmatrix} 0 && 0 && \sigma^{x}_{xz}\\0 && 0 && -\sigma^{y}_{xz}\\\sigma^{x}_{zx} && -\sigma^{y}_{zx} && 0 \end{pmatrix}\end{math} & \begin{math} \begin{pmatrix} 0 && 0 && \sigma^{y}_{xz}\\0 && 0 && \sigma^{x}_{xz}\\\sigma^{y}_{zx} && \sigma^{x}_{zx} && 0 \end{pmatrix}\end{math} & \begin{math} \begin{pmatrix} \sigma^{z}_{xx} && \sigma^{z}_{xy} && 0\\-\sigma^{z}_{xy} && \sigma^{z}_{xx} && 0\\0 && 0 && \sigma^{z}_{zz} \end{pmatrix}\end{math}\\[4mm]
 \spacegroup{4/mmm1'}, \spacegroup{6/mmm1'} & \begin{math} \begin{pmatrix} 0 && 0 && 0\\0 && 0 && -\sigma^{y}_{xz}\\0 && -\sigma^{z}_{xy} && 0 \end{pmatrix}\end{math} & \begin{math} \begin{pmatrix} 0 && 0 && \sigma^{y}_{xz}\\0 && 0 && 0\\\sigma^{z}_{xy} && 0 && 0 \end{pmatrix}\end{math} & \begin{math} \begin{pmatrix} 0 && \sigma^{y}_{zx} && 0\\-\sigma^{y}_{zx} && 0 && 0\\0 && 0 && 0 \end{pmatrix}\end{math} & \begin{math} \begin{pmatrix} 0 && 0 && 0\\0 && 0 && -\sigma^{y}_{xz}\\0 && -\sigma^{y}_{zx} && 0 \end{pmatrix}\end{math} & \begin{math} \begin{pmatrix} 0 && 0 && \sigma^{y}_{xz}\\0 && 0 && 0\\\sigma^{y}_{zx} && 0 && 0 \end{pmatrix}\end{math} & \begin{math} \begin{pmatrix} 0 && \sigma^{z}_{xy} && 0\\-\sigma^{z}_{xy} && 0 && 0\\0 && 0 && 0 \end{pmatrix}\end{math}\\[4mm]
 \spacegroup{\bar 31'} & \begin{math} \begin{pmatrix} -\sigma^{x}_{xx} && -\sigma^{y}_{xx} && -\sigma^{x}_{xz}\\-\sigma^{y}_{xx} && \sigma^{x}_{xx} && -\sigma^{y}_{xz}\\-\sigma^{z}_{xx} && -\sigma^{z}_{xy} && 0 \end{pmatrix}\end{math} & \begin{math} \begin{pmatrix} -\sigma^{y}_{xx} && \sigma^{x}_{xx} && \sigma^{y}_{xz}\\\sigma^{x}_{xx} && \sigma^{y}_{xx} && -\sigma^{x}_{xz}\\\sigma^{z}_{xy} && -\sigma^{z}_{xx} && 0 \end{pmatrix}\end{math} & \begin{math} \begin{pmatrix} -\sigma^{x}_{zx} && \sigma^{y}_{zx} && 0\\-\sigma^{y}_{zx} && -\sigma^{x}_{zx} && 0\\0 && 0 && -\sigma^{z}_{zz} \end{pmatrix}\end{math} & \begin{math} \begin{pmatrix} \sigma^{x}_{xx} && \sigma^{y}_{xx} && \sigma^{x}_{xz}\\\sigma^{y}_{xx} && -\sigma^{x}_{xx} && -\sigma^{y}_{xz}\\\sigma^{x}_{zx} && -\sigma^{y}_{zx} && 0 \end{pmatrix}\end{math} & \begin{math} \begin{pmatrix} \sigma^{y}_{xx} && -\sigma^{x}_{xx} && \sigma^{y}_{xz}\\-\sigma^{x}_{xx} && -\sigma^{y}_{xx} && \sigma^{x}_{xz}\\\sigma^{y}_{zx} && \sigma^{x}_{zx} && 0 \end{pmatrix}\end{math} & \begin{math} \begin{pmatrix} \sigma^{z}_{xx} && \sigma^{z}_{xy} && 0\\-\sigma^{z}_{xy} && \sigma^{z}_{xx} && 0\\0 && 0 && \sigma^{z}_{zz} \end{pmatrix}\end{math}\\[4mm]
 \spacegroup{\bar 31m1'} & \begin{math} \begin{pmatrix} -\sigma^{x}_{xx} && 0 && 0\\0 && \sigma^{x}_{xx} && -\sigma^{y}_{xz}\\0 && -\sigma^{z}_{xy} && 0 \end{pmatrix}\end{math} & \begin{math} \begin{pmatrix} 0 && \sigma^{x}_{xx} && \sigma^{y}_{xz}\\\sigma^{x}_{xx} && 0 && 0\\\sigma^{z}_{xy} && 0 && 0 \end{pmatrix}\end{math} & \begin{math} \begin{pmatrix} 0 && \sigma^{y}_{zx} && 0\\-\sigma^{y}_{zx} && 0 && 0\\0 && 0 && 0 \end{pmatrix}\end{math} & \begin{math} \begin{pmatrix} \sigma^{x}_{xx} && 0 && 0\\0 && -\sigma^{x}_{xx} && -\sigma^{y}_{xz}\\0 && -\sigma^{y}_{zx} && 0 \end{pmatrix}\end{math} & \begin{math} \begin{pmatrix} 0 && -\sigma^{x}_{xx} && \sigma^{y}_{xz}\\-\sigma^{x}_{xx} && 0 && 0\\\sigma^{y}_{zx} && 0 && 0 \end{pmatrix}\end{math} & \begin{math} \begin{pmatrix} 0 && \sigma^{z}_{xy} && 0\\-\sigma^{z}_{xy} && 0 && 0\\0 && 0 && 0 \end{pmatrix}\end{math}\\[4mm]
 \spacegroup{\bar 3m11'} & \begin{math} \begin{pmatrix} 0 && -\sigma^{y}_{xx} && 0\\-\sigma^{y}_{xx} && 0 && -\sigma^{y}_{xz}\\0 && -\sigma^{z}_{xy} && 0 \end{pmatrix}\end{math} & \begin{math} \begin{pmatrix} -\sigma^{y}_{xx} && 0 && \sigma^{y}_{xz}\\0 && \sigma^{y}_{xx} && 0\\\sigma^{z}_{xy} && 0 && 0 \end{pmatrix}\end{math} & \begin{math} \begin{pmatrix} 0 && \sigma^{y}_{zx} && 0\\-\sigma^{y}_{zx} && 0 && 0\\0 && 0 && 0 \end{pmatrix}\end{math} & \begin{math} \begin{pmatrix} 0 && \sigma^{y}_{xx} && 0\\\sigma^{y}_{xx} && 0 && -\sigma^{y}_{xz}\\0 && -\sigma^{y}_{zx} && 0 \end{pmatrix}\end{math} & \begin{math} \begin{pmatrix} \sigma^{y}_{xx} && 0 && \sigma^{y}_{xz}\\0 && -\sigma^{y}_{xx} && 0\\\sigma^{y}_{zx} && 0 && 0 \end{pmatrix}\end{math} & \begin{math} \begin{pmatrix} 0 && \sigma^{z}_{xy} && 0\\-\sigma^{z}_{xy} && 0 && 0\\0 && 0 && 0 \end{pmatrix}\end{math}\\[4mm]
 \spacegroup{m\bar 31'} & \begin{math} \begin{pmatrix} 0 && 0 && 0\\0 && 0 && -\sigma^{y}_{xz}\\0 && -\sigma^{z}_{xy} && 0 \end{pmatrix}\end{math} & \begin{math} \begin{pmatrix} 0 && 0 && -\sigma^{z}_{xy}\\0 && 0 && 0\\-\sigma^{y}_{xz} && 0 && 0 \end{pmatrix}\end{math} & \begin{math} \begin{pmatrix} 0 && -\sigma^{y}_{xz} && 0\\-\sigma^{z}_{xy} && 0 && 0\\0 && 0 && 0 \end{pmatrix}\end{math} & \begin{math} \begin{pmatrix} 0 && 0 && 0\\0 && 0 && \sigma^{z}_{xy}\\0 && \sigma^{y}_{xz} && 0 \end{pmatrix}\end{math} & \begin{math} \begin{pmatrix} 0 && 0 && \sigma^{y}_{xz}\\0 && 0 && 0\\\sigma^{z}_{xy} && 0 && 0 \end{pmatrix}\end{math} & \begin{math} \begin{pmatrix} 0 && \sigma^{z}_{xy} && 0\\\sigma^{y}_{xz} && 0 && 0\\0 && 0 && 0 \end{pmatrix}\end{math}\\[4mm]
 \spacegroup{m\bar 3m1'} & \begin{math} \begin{pmatrix} 0 && 0 && 0\\0 && 0 && \sigma^{z}_{xy}\\0 && -\sigma^{z}_{xy} && 0 \end{pmatrix}\end{math} & \begin{math} \begin{pmatrix} 0 && 0 && -\sigma^{z}_{xy}\\0 && 0 && 0\\\sigma^{z}_{xy} && 0 && 0 \end{pmatrix}\end{math} & \begin{math} \begin{pmatrix} 0 && \sigma^{z}_{xy} && 0\\-\sigma^{z}_{xy} && 0 && 0\\0 && 0 && 0 \end{pmatrix}\end{math} & \begin{math} \begin{pmatrix} 0 && 0 && 0\\0 && 0 && \sigma^{z}_{xy}\\0 && -\sigma^{z}_{xy} && 0 \end{pmatrix}\end{math} & \begin{math} \begin{pmatrix} 0 && 0 && -\sigma^{z}_{xy}\\0 && 0 && 0\\\sigma^{z}_{xy} && 0 && 0 \end{pmatrix}\end{math} & \begin{math} \begin{pmatrix} 0 && \sigma^{z}_{xy} && 0\\-\sigma^{z}_{xy} && 0 && 0\\0 && 0 && 0 \end{pmatrix}\end{math}\\[4mm]
 \bottomrule
\end{tabular}

%% file: tensor_pol_and_reduced_cat_b.tex
\begin{tabular}{>{\centering}m{1.5cm}| ccc|ccc}
\toprule \tableheading{magnetic \name{Laue} group} & \tableheading{$\matrg \sigma'^x$} & \tableheading{$\matrg \sigma'^y$} & \tableheading{$\matrg \sigma'^z$} & \tableheading{$\matrg \sigma^x$} & \tableheading{$\matrg \sigma^y$} & \tableheading{$\matrg \sigma^z$} \\[4mm]
 \midrule
\spacegroup{\bar 1} & \begin{math} \begin{pmatrix} \sigma^{x}_{xx} && \sigma^{x}_{xy} && \sigma^{x}_{xz}\\\sigma^{x}_{yx} && \sigma^{x}_{yy} && \sigma^{x}_{yz}\\\sigma^{x}_{zx} && \sigma^{x}_{zy} && \sigma^{x}_{zz} \end{pmatrix}\end{math} & \begin{math} \begin{pmatrix} \sigma^{y}_{xx} && \sigma^{y}_{xy} && \sigma^{y}_{xz}\\\sigma^{y}_{yx} && \sigma^{y}_{yy} && \sigma^{y}_{yz}\\\sigma^{y}_{zx} && \sigma^{y}_{zy} && \sigma^{y}_{zz} \end{pmatrix}\end{math} & \begin{math} \begin{pmatrix} \sigma^{z}_{xx} && \sigma^{z}_{xy} && \sigma^{z}_{xz}\\\sigma^{z}_{yx} && \sigma^{z}_{yy} && \sigma^{z}_{yz}\\\sigma^{z}_{zx} && \sigma^{z}_{zy} && \sigma^{z}_{zz} \end{pmatrix}\end{math} & \begin{math} \begin{pmatrix} \sigma^{x}_{xx} && \sigma^{x}_{xy} && \sigma^{x}_{xz}\\\sigma^{x}_{yx} && \sigma^{x}_{yy} && \sigma^{x}_{yz}\\\sigma^{x}_{zx} && \sigma^{x}_{zy} && \sigma^{x}_{zz} \end{pmatrix}\end{math} & \begin{math} \begin{pmatrix} \sigma^{y}_{xx} && \sigma^{y}_{xy} && \sigma^{y}_{xz}\\\sigma^{y}_{yx} && \sigma^{y}_{yy} && \sigma^{y}_{yz}\\\sigma^{y}_{zx} && \sigma^{y}_{zy} && \sigma^{y}_{zz} \end{pmatrix}\end{math} & \begin{math} \begin{pmatrix} \sigma^{z}_{xx} && \sigma^{z}_{xy} && \sigma^{z}_{xz}\\\sigma^{z}_{yx} && \sigma^{z}_{yy} && \sigma^{z}_{yz}\\\sigma^{z}_{zx} && \sigma^{z}_{zy} && \sigma^{z}_{zz} \end{pmatrix}\end{math}\\[4mm]
 \spacegroup{2/m} & \begin{math} \begin{pmatrix} 0 && \sigma^{x}_{xy} && 0\\\sigma^{x}_{yx} && 0 && \sigma^{x}_{yz}\\0 && \sigma^{x}_{zy} && 0 \end{pmatrix}\end{math} & \begin{math} \begin{pmatrix} \sigma^{y}_{xx} && 0 && \sigma^{y}_{xz}\\0 && \sigma^{y}_{yy} && 0\\\sigma^{y}_{zx} && 0 && \sigma^{y}_{zz} \end{pmatrix}\end{math} & \begin{math} \begin{pmatrix} 0 && \sigma^{z}_{xy} && 0\\\sigma^{z}_{yx} && 0 && \sigma^{z}_{yz}\\0 && \sigma^{z}_{zy} && 0 \end{pmatrix}\end{math} & \begin{math} \begin{pmatrix} 0 && \sigma^{x}_{xy} && 0\\\sigma^{x}_{yx} && 0 && \sigma^{x}_{yz}\\0 && \sigma^{x}_{zy} && 0 \end{pmatrix}\end{math} & \begin{math} \begin{pmatrix} \sigma^{y}_{xx} && 0 && \sigma^{y}_{xz}\\0 && \sigma^{y}_{yy} && 0\\\sigma^{y}_{zx} && 0 && \sigma^{y}_{zz} \end{pmatrix}\end{math} & \begin{math} \begin{pmatrix} 0 && \sigma^{z}_{xy} && 0\\\sigma^{z}_{yx} && 0 && \sigma^{z}_{yz}\\0 && \sigma^{z}_{zy} && 0 \end{pmatrix}\end{math}\\[4mm]
 \spacegroup{mmm} & \begin{math} \begin{pmatrix} 0 && 0 && 0\\0 && 0 && \sigma^{x}_{yz}\\0 && \sigma^{x}_{zy} && 0 \end{pmatrix}\end{math} & \begin{math} \begin{pmatrix} 0 && 0 && \sigma^{y}_{xz}\\0 && 0 && 0\\\sigma^{y}_{zx} && 0 && 0 \end{pmatrix}\end{math} & \begin{math} \begin{pmatrix} 0 && \sigma^{z}_{xy} && 0\\\sigma^{z}_{yx} && 0 && 0\\0 && 0 && 0 \end{pmatrix}\end{math} & \begin{math} \begin{pmatrix} 0 && 0 && 0\\0 && 0 && \sigma^{x}_{yz}\\0 && \sigma^{x}_{zy} && 0 \end{pmatrix}\end{math} & \begin{math} \begin{pmatrix} 0 && 0 && \sigma^{y}_{xz}\\0 && 0 && 0\\\sigma^{y}_{zx} && 0 && 0 \end{pmatrix}\end{math} & \begin{math} \begin{pmatrix} 0 && \sigma^{z}_{xy} && 0\\\sigma^{z}_{yx} && 0 && 0\\0 && 0 && 0 \end{pmatrix}\end{math}\\[4mm]
 \spacegroup{4/m}, \spacegroup{6/m} & \begin{math} \begin{pmatrix} 0 && 0 && \sigma^{x}_{xz}\\0 && 0 && -\sigma^{y}_{xz}\\\sigma^{x}_{zx} && -\sigma^{y}_{zx} && 0 \end{pmatrix}\end{math} & \begin{math} \begin{pmatrix} 0 && 0 && \sigma^{y}_{xz}\\0 && 0 && \sigma^{x}_{xz}\\\sigma^{y}_{zx} && \sigma^{x}_{zx} && 0 \end{pmatrix}\end{math} & \begin{math} \begin{pmatrix} \sigma^{z}_{xx} && \sigma^{z}_{xy} && 0\\-\sigma^{z}_{xy} && \sigma^{z}_{xx} && 0\\0 && 0 && \sigma^{z}_{zz} \end{pmatrix}\end{math} & \begin{math} \begin{pmatrix} 0 && 0 && \sigma^{x}_{xz}\\0 && 0 && -\sigma^{y}_{xz}\\\sigma^{x}_{zx} && -\sigma^{y}_{zx} && 0 \end{pmatrix}\end{math} & \begin{math} \begin{pmatrix} 0 && 0 && \sigma^{y}_{xz}\\0 && 0 && \sigma^{x}_{xz}\\\sigma^{y}_{zx} && \sigma^{x}_{zx} && 0 \end{pmatrix}\end{math} & \begin{math} \begin{pmatrix} \sigma^{z}_{xx} && \sigma^{z}_{xy} && 0\\-\sigma^{z}_{xy} && \sigma^{z}_{xx} && 0\\0 && 0 && \sigma^{z}_{zz} \end{pmatrix}\end{math}\\[4mm]
 \spacegroup{4/mmm}, \spacegroup{6/mmm} & \begin{math} \begin{pmatrix} 0 && 0 && 0\\0 && 0 && -\sigma^{y}_{xz}\\0 && -\sigma^{y}_{zx} && 0 \end{pmatrix}\end{math} & \begin{math} \begin{pmatrix} 0 && 0 && \sigma^{y}_{xz}\\0 && 0 && 0\\\sigma^{y}_{zx} && 0 && 0 \end{pmatrix}\end{math} & \begin{math} \begin{pmatrix} 0 && \sigma^{z}_{xy} && 0\\-\sigma^{z}_{xy} && 0 && 0\\0 && 0 && 0 \end{pmatrix}\end{math} & \begin{math} \begin{pmatrix} 0 && 0 && 0\\0 && 0 && -\sigma^{y}_{xz}\\0 && -\sigma^{y}_{zx} && 0 \end{pmatrix}\end{math} & \begin{math} \begin{pmatrix} 0 && 0 && \sigma^{y}_{xz}\\0 && 0 && 0\\\sigma^{y}_{zx} && 0 && 0 \end{pmatrix}\end{math} & \begin{math} \begin{pmatrix} 0 && \sigma^{z}_{xy} && 0\\-\sigma^{z}_{xy} && 0 && 0\\0 && 0 && 0 \end{pmatrix}\end{math}\\[4mm]
 \spacegroup{\bar 3} & \begin{math} \begin{pmatrix} \sigma^{x}_{xx} && \sigma^{y}_{xx} && \sigma^{x}_{xz}\\\sigma^{y}_{xx} && -\sigma^{x}_{xx} && -\sigma^{y}_{xz}\\\sigma^{x}_{zx} && -\sigma^{y}_{zx} && 0 \end{pmatrix}\end{math} & \begin{math} \begin{pmatrix} \sigma^{y}_{xx} && -\sigma^{x}_{xx} && \sigma^{y}_{xz}\\-\sigma^{x}_{xx} && -\sigma^{y}_{xx} && \sigma^{x}_{xz}\\\sigma^{y}_{zx} && \sigma^{x}_{zx} && 0 \end{pmatrix}\end{math} & \begin{math} \begin{pmatrix} \sigma^{z}_{xx} && \sigma^{z}_{xy} && 0\\-\sigma^{z}_{xy} && \sigma^{z}_{xx} && 0\\0 && 0 && \sigma^{z}_{zz} \end{pmatrix}\end{math} & \begin{math} \begin{pmatrix} \sigma^{x}_{xx} && \sigma^{y}_{xx} && \sigma^{x}_{xz}\\\sigma^{y}_{xx} && -\sigma^{x}_{xx} && -\sigma^{y}_{xz}\\\sigma^{x}_{zx} && -\sigma^{y}_{zx} && 0 \end{pmatrix}\end{math} & \begin{math} \begin{pmatrix} \sigma^{y}_{xx} && -\sigma^{x}_{xx} && \sigma^{y}_{xz}\\-\sigma^{x}_{xx} && -\sigma^{y}_{xx} && \sigma^{x}_{xz}\\\sigma^{y}_{zx} && \sigma^{x}_{zx} && 0 \end{pmatrix}\end{math} & \begin{math} \begin{pmatrix} \sigma^{z}_{xx} && \sigma^{z}_{xy} && 0\\-\sigma^{z}_{xy} && \sigma^{z}_{xx} && 0\\0 && 0 && \sigma^{z}_{zz} \end{pmatrix}\end{math}\\[4mm]
 \spacegroup{\bar 31m} & \begin{math} \begin{pmatrix} \sigma^{x}_{xx} && 0 && 0\\0 && -\sigma^{x}_{xx} && -\sigma^{y}_{xz}\\0 && -\sigma^{y}_{zx} && 0 \end{pmatrix}\end{math} & \begin{math} \begin{pmatrix} 0 && -\sigma^{x}_{xx} && \sigma^{y}_{xz}\\-\sigma^{x}_{xx} && 0 && 0\\\sigma^{y}_{zx} && 0 && 0 \end{pmatrix}\end{math} & \begin{math} \begin{pmatrix} 0 && \sigma^{z}_{xy} && 0\\-\sigma^{z}_{xy} && 0 && 0\\0 && 0 && 0 \end{pmatrix}\end{math} & \begin{math} \begin{pmatrix} \sigma^{x}_{xx} && 0 && 0\\0 && -\sigma^{x}_{xx} && -\sigma^{y}_{xz}\\0 && -\sigma^{y}_{zx} && 0 \end{pmatrix}\end{math} & \begin{math} \begin{pmatrix} 0 && -\sigma^{x}_{xx} && \sigma^{y}_{xz}\\-\sigma^{x}_{xx} && 0 && 0\\\sigma^{y}_{zx} && 0 && 0 \end{pmatrix}\end{math} & \begin{math} \begin{pmatrix} 0 && \sigma^{z}_{xy} && 0\\-\sigma^{z}_{xy} && 0 && 0\\0 && 0 && 0 \end{pmatrix}\end{math}\\[4mm]
 \spacegroup{\bar 3m1} & \begin{math} \begin{pmatrix} 0 && \sigma^{y}_{xx} && 0\\\sigma^{y}_{xx} && 0 && -\sigma^{y}_{xz}\\0 && -\sigma^{y}_{zx} && 0 \end{pmatrix}\end{math} & \begin{math} \begin{pmatrix} \sigma^{y}_{xx} && 0 && \sigma^{y}_{xz}\\0 && -\sigma^{y}_{xx} && 0\\\sigma^{y}_{zx} && 0 && 0 \end{pmatrix}\end{math} & \begin{math} \begin{pmatrix} 0 && \sigma^{z}_{xy} && 0\\-\sigma^{z}_{xy} && 0 && 0\\0 && 0 && 0 \end{pmatrix}\end{math} & \begin{math} \begin{pmatrix} 0 && \sigma^{y}_{xx} && 0\\\sigma^{y}_{xx} && 0 && -\sigma^{y}_{xz}\\0 && -\sigma^{y}_{zx} && 0 \end{pmatrix}\end{math} & \begin{math} \begin{pmatrix} \sigma^{y}_{xx} && 0 && \sigma^{y}_{xz}\\0 && -\sigma^{y}_{xx} && 0\\\sigma^{y}_{zx} && 0 && 0 \end{pmatrix}\end{math} & \begin{math} \begin{pmatrix} 0 && \sigma^{z}_{xy} && 0\\-\sigma^{z}_{xy} && 0 && 0\\0 && 0 && 0 \end{pmatrix}\end{math}\\[4mm]
 \spacegroup{m\bar 3} & \begin{math} \begin{pmatrix} 0 && 0 && 0\\0 && 0 && \sigma^{z}_{xy}\\0 && \sigma^{y}_{xz} && 0 \end{pmatrix}\end{math} & \begin{math} \begin{pmatrix} 0 && 0 && \sigma^{y}_{xz}\\0 && 0 && 0\\\sigma^{z}_{xy} && 0 && 0 \end{pmatrix}\end{math} & \begin{math} \begin{pmatrix} 0 && \sigma^{z}_{xy} && 0\\\sigma^{y}_{xz} && 0 && 0\\0 && 0 && 0 \end{pmatrix}\end{math} & \begin{math} \begin{pmatrix} 0 && 0 && 0\\0 && 0 && \sigma^{z}_{xy}\\0 && \sigma^{y}_{xz} && 0 \end{pmatrix}\end{math} & \begin{math} \begin{pmatrix} 0 && 0 && \sigma^{y}_{xz}\\0 && 0 && 0\\\sigma^{z}_{xy} && 0 && 0 \end{pmatrix}\end{math} & \begin{math} \begin{pmatrix} 0 && \sigma^{z}_{xy} && 0\\\sigma^{y}_{xz} && 0 && 0\\0 && 0 && 0 \end{pmatrix}\end{math}\\[4mm]
 \spacegroup{m\bar 3m} & \begin{math} \begin{pmatrix} 0 && 0 && 0\\0 && 0 && \sigma^{z}_{xy}\\0 && -\sigma^{z}_{xy} && 0 \end{pmatrix}\end{math} & \begin{math} \begin{pmatrix} 0 && 0 && -\sigma^{z}_{xy}\\0 && 0 && 0\\\sigma^{z}_{xy} && 0 && 0 \end{pmatrix}\end{math} & \begin{math} \begin{pmatrix} 0 && \sigma^{z}_{xy} && 0\\-\sigma^{z}_{xy} && 0 && 0\\0 && 0 && 0 \end{pmatrix}\end{math} & \begin{math} \begin{pmatrix} 0 && 0 && 0\\0 && 0 && \sigma^{z}_{xy}\\0 && -\sigma^{z}_{xy} && 0 \end{pmatrix}\end{math} & \begin{math} \begin{pmatrix} 0 && 0 && -\sigma^{z}_{xy}\\0 && 0 && 0\\\sigma^{z}_{xy} && 0 && 0 \end{pmatrix}\end{math} & \begin{math} \begin{pmatrix} 0 && \sigma^{z}_{xy} && 0\\-\sigma^{z}_{xy} && 0 && 0\\0 && 0 && 0 \end{pmatrix}\end{math}\\[4mm]
 \bottomrule
\end{tabular}

%% file: tensor_pol_and_reduced_cat_c_new.tex
\begin{tabular}{>{\centering}m{1.5cm}| ccc| ccc}
\toprule \tableheading{magnetic \name{Laue} group} & \tableheading{$\matrg \sigma^{\prime x}$} & \tableheading{$\matrg \sigma^{\prime y}$} & \tableheading{$\matrg \sigma^{\prime z}$} & \tableheading{$\matrg \sigma^{x}$} & \tableheading{$\matrg \sigma^{y}$} & \tableheading{$\matrg \sigma^{z}$} \\[4mm]
 \midrule
\spacegroup{2'/m'} & \begin{math} \begin{pmatrix} \sigma^{x}_{xx} && -\sigma^{x}_{xy} && \sigma^{x}_{xz}\\-\sigma^{y}_{xx} && \sigma^{y}_{xy} && -\sigma^{y}_{xz}\\\sigma^{z}_{xx} && -\sigma^{z}_{xy} && \sigma^{z}_{xz} \end{pmatrix}\end{math} & \begin{math} \begin{pmatrix} -\sigma^{x}_{yx} && \sigma^{x}_{yy} && -\sigma^{x}_{yz}\\\sigma^{y}_{yx} && -\sigma^{y}_{yy} && \sigma^{y}_{yz}\\-\sigma^{z}_{yx} && \sigma^{z}_{yy} && -\sigma^{z}_{yz} \end{pmatrix}\end{math} & \begin{math} \begin{pmatrix} \sigma^{x}_{zx} && -\sigma^{x}_{zy} && \sigma^{x}_{zz}\\-\sigma^{y}_{zx} && \sigma^{y}_{zy} && -\sigma^{y}_{zz}\\\sigma^{z}_{zx} && -\sigma^{z}_{zy} && \sigma^{z}_{zz} \end{pmatrix}\end{math} &  \begin{math} \begin{pmatrix} \sigma^{x}_{xx} && \sigma^{x}_{xy} && \sigma^{x}_{xz}\\\sigma^{x}_{yx} && \sigma^{x}_{yy} && \sigma^{x}_{yz}\\\sigma^{x}_{zx} && \sigma^{x}_{zy} && \sigma^{x}_{zz} \end{pmatrix}\end{math} & \begin{math} \begin{pmatrix} \sigma^{y}_{xx} && \sigma^{y}_{xy} && \sigma^{y}_{xz}\\\sigma^{y}_{yx} && \sigma^{y}_{yy} && \sigma^{y}_{yz}\\\sigma^{y}_{zx} && \sigma^{y}_{zy} && \sigma^{y}_{zz} \end{pmatrix}\end{math} & \begin{math} \begin{pmatrix} \sigma^{z}_{xx} && \sigma^{z}_{xy} && \sigma^{z}_{xz}\\\sigma^{z}_{yx} && \sigma^{z}_{yy} && \sigma^{z}_{yz}\\\sigma^{z}_{zx} && \sigma^{z}_{zy} && \sigma^{z}_{zz} \end{pmatrix}\end{math}\\[4mm]
 \spacegroup{m'm'm} & \begin{math} \begin{pmatrix} 0 && 0 && \sigma^{x}_{xz}\\0 && 0 && -\sigma^{y}_{xz}\\\sigma^{z}_{xx} && -\sigma^{z}_{xy} && 0 \end{pmatrix}\end{math} & \begin{math} \begin{pmatrix} 0 && 0 && -\sigma^{x}_{yz}\\0 && 0 && \sigma^{y}_{yz}\\-\sigma^{z}_{yx} && \sigma^{z}_{yy} && 0 \end{pmatrix}\end{math} & \begin{math} \begin{pmatrix} \sigma^{x}_{zx} && -\sigma^{x}_{zy} && 0\\-\sigma^{y}_{zx} && \sigma^{y}_{zy} && 0\\0 && 0 && \sigma^{z}_{zz} \end{pmatrix}\end{math} & \begin{math} \begin{pmatrix} 0 && 0 && \sigma^{x}_{xz}\\0 && 0 && \sigma^{x}_{yz}\\\sigma^{x}_{zx} && \sigma^{x}_{zy} && 0 \end{pmatrix}\end{math} & \begin{math} \begin{pmatrix} 0 && 0 && \sigma^{y}_{xz}\\0 && 0 && \sigma^{y}_{yz}\\\sigma^{y}_{zx} && \sigma^{y}_{zy} && 0 \end{pmatrix}\end{math} & \begin{math} \begin{pmatrix} \sigma^{z}_{xx} && \sigma^{z}_{xy} && 0\\\sigma^{z}_{yx} && \sigma^{z}_{yy} && 0\\0 && 0 && \sigma^{z}_{zz} \end{pmatrix}\end{math}\\[4mm]
 \spacegroup{4'/m} & \begin{math} \begin{pmatrix} 0 && 0 && -\sigma^{y}_{yz}\\0 && 0 && \sigma^{x}_{yz}\\-\sigma^{z}_{yy} && \sigma^{z}_{yx} && 0 \end{pmatrix}\end{math} & \begin{math} \begin{pmatrix} 0 && 0 && \sigma^{y}_{xz}\\0 && 0 && -\sigma^{x}_{xz}\\\sigma^{z}_{xy} && -\sigma^{z}_{xx} && 0 \end{pmatrix}\end{math} & \begin{math} \begin{pmatrix} -\sigma^{y}_{zy} && \sigma^{y}_{zx} && 0\\\sigma^{x}_{zy} && -\sigma^{x}_{zx} && 0\\0 && 0 && -\sigma^{z}_{zz} \end{pmatrix}\end{math} & \begin{math} \begin{pmatrix} 0 && 0 && \sigma^{x}_{xz}\\0 && 0 && \sigma^{x}_{yz}\\\sigma^{x}_{zx} && \sigma^{x}_{zy} && 0 \end{pmatrix}\end{math} & \begin{math} \begin{pmatrix} 0 && 0 && \sigma^{y}_{xz}\\0 && 0 && \sigma^{y}_{yz}\\\sigma^{y}_{zx} && \sigma^{y}_{zy} && 0 \end{pmatrix}\end{math} & \begin{math} \begin{pmatrix} \sigma^{z}_{xx} && \sigma^{z}_{xy} && 0\\\sigma^{z}_{yx} && \sigma^{z}_{yy} && 0\\0 && 0 && \sigma^{z}_{zz} \end{pmatrix}\end{math}\\[4mm]
 \spacegroup{4'/mm'm} & \begin{math} \begin{pmatrix} 0 && 0 && \sigma^{x}_{xz}\\0 && 0 && -\sigma^{y}_{xz}\\\sigma^{z}_{xx} && -\sigma^{z}_{xy} && 0 \end{pmatrix}\end{math} & \begin{math} \begin{pmatrix} 0 && 0 && \sigma^{y}_{xz}\\0 && 0 && -\sigma^{x}_{xz}\\\sigma^{z}_{xy} && -\sigma^{z}_{xx} && 0 \end{pmatrix}\end{math} & \begin{math} \begin{pmatrix} \sigma^{x}_{zx} && \sigma^{y}_{zx} && 0\\-\sigma^{y}_{zx} && -\sigma^{x}_{zx} && 0\\0 && 0 && 0 \end{pmatrix}\end{math} & \begin{math} \begin{pmatrix} 0 && 0 && \sigma^{x}_{xz}\\0 && 0 && -\sigma^{y}_{xz}\\\sigma^{x}_{zx} && -\sigma^{y}_{zx} && 0 \end{pmatrix}\end{math} & \begin{math} \begin{pmatrix} 0 && 0 && \sigma^{y}_{xz}\\0 && 0 && -\sigma^{x}_{xz}\\\sigma^{y}_{zx} && -\sigma^{x}_{zx} && 0 \end{pmatrix}\end{math} & \begin{math} \begin{pmatrix} \sigma^{z}_{xx} && \sigma^{z}_{xy} && 0\\-\sigma^{z}_{xy} && -\sigma^{z}_{xx} && 0\\0 && 0 && 0 \end{pmatrix}\end{math}\\[4mm]
 \spacegroup{4'/mmm'} & \begin{math} \begin{pmatrix} 0 && 0 && 0\\0 && 0 && \sigma^{x}_{yz}\\0 && \sigma^{z}_{yx} && 0 \end{pmatrix}\end{math} & \begin{math} \begin{pmatrix} 0 && 0 && \sigma^{y}_{xz}\\0 && 0 && 0\\\sigma^{z}_{xy} && 0 && 0 \end{pmatrix}\end{math} & \begin{math} \begin{pmatrix} 0 && \sigma^{y}_{zx} && 0\\\sigma^{x}_{zy} && 0 && 0\\0 && 0 && 0 \end{pmatrix}\end{math} & \begin{math} \begin{pmatrix} 0 && 0 && 0\\0 && 0 && \sigma^{x}_{yz}\\0 && \sigma^{x}_{zy} && 0 \end{pmatrix}\end{math} & \begin{math} \begin{pmatrix} 0 && 0 && \sigma^{y}_{xz}\\0 && 0 && 0\\\sigma^{y}_{zx} && 0 && 0 \end{pmatrix}\end{math} & \begin{math} \begin{pmatrix} 0 && \sigma^{z}_{xy} && 0\\\sigma^{z}_{yx} && 0 && 0\\0 && 0 && 0 \end{pmatrix}\end{math}\\[4mm]
 \spacegroup{4/mm'm'}, \spacegroup{6/mm'm'} & \begin{math} \begin{pmatrix} 0 && 0 && \sigma^{x}_{xz}\\0 && 0 && -\sigma^{y}_{xz}\\\sigma^{z}_{xx} && -\sigma^{z}_{xy} && 0 \end{pmatrix}\end{math} & \begin{math} \begin{pmatrix} 0 && 0 && \sigma^{y}_{xz}\\0 && 0 && \sigma^{x}_{xz}\\\sigma^{z}_{xy} && \sigma^{z}_{xx} && 0 \end{pmatrix}\end{math} & \begin{math} \begin{pmatrix} \sigma^{x}_{zx} && \sigma^{y}_{zx} && 0\\-\sigma^{y}_{zx} && \sigma^{x}_{zx} && 0\\0 && 0 && \sigma^{z}_{zz} \end{pmatrix}\end{math} & \begin{math} \begin{pmatrix} 0 && 0 && \sigma^{x}_{xz}\\0 && 0 && -\sigma^{y}_{xz}\\\sigma^{x}_{zx} && -\sigma^{y}_{zx} && 0 \end{pmatrix}\end{math} & \begin{math} \begin{pmatrix} 0 && 0 && \sigma^{y}_{xz}\\0 && 0 && \sigma^{x}_{xz}\\\sigma^{y}_{zx} && \sigma^{x}_{zx} && 0 \end{pmatrix}\end{math} & \begin{math} \begin{pmatrix} \sigma^{z}_{xx} && \sigma^{z}_{xy} && 0\\-\sigma^{z}_{xy} && \sigma^{z}_{xx} && 0\\0 && 0 && \sigma^{z}_{zz} \end{pmatrix}\end{math}\\[4mm]
 \spacegroup{\bar 31m'} & \begin{math} \begin{pmatrix} -\sigma^{x}_{xx} && \sigma^{y}_{xx} && \sigma^{x}_{xz}\\\sigma^{y}_{xx} && \sigma^{x}_{xx} && -\sigma^{y}_{xz}\\\sigma^{z}_{xx} && -\sigma^{z}_{xy} && 0 \end{pmatrix}\end{math} & \begin{math} \begin{pmatrix} \sigma^{y}_{xx} && \sigma^{x}_{xx} && \sigma^{y}_{xz}\\\sigma^{x}_{xx} && -\sigma^{y}_{xx} && \sigma^{x}_{xz}\\\sigma^{z}_{xy} && \sigma^{z}_{xx} && 0 \end{pmatrix}\end{math} & \begin{math} \begin{pmatrix} \sigma^{x}_{zx} && \sigma^{y}_{zx} && 0\\-\sigma^{y}_{zx} && \sigma^{x}_{zx} && 0\\0 && 0 && \sigma^{z}_{zz} \end{pmatrix}\end{math} & \begin{math} \begin{pmatrix} \sigma^{x}_{xx} && \sigma^{y}_{xx} && \sigma^{x}_{xz}\\\sigma^{y}_{xx} && -\sigma^{x}_{xx} && -\sigma^{y}_{xz}\\\sigma^{x}_{zx} && -\sigma^{y}_{zx} && 0 \end{pmatrix}\end{math} & \begin{math} \begin{pmatrix} \sigma^{y}_{xx} && -\sigma^{x}_{xx} && \sigma^{y}_{xz}\\-\sigma^{x}_{xx} && -\sigma^{y}_{xx} && \sigma^{x}_{xz}\\\sigma^{y}_{zx} && \sigma^{x}_{zx} && 0 \end{pmatrix}\end{math} & \begin{math} \begin{pmatrix} \sigma^{z}_{xx} && \sigma^{z}_{xy} && 0\\-\sigma^{z}_{xy} && \sigma^{z}_{xx} && 0\\0 && 0 && \sigma^{z}_{zz} \end{pmatrix}\end{math}\\[4mm]
 \spacegroup{\bar 3m'1} & \begin{math} \begin{pmatrix} \sigma^{x}_{xx} && -\sigma^{y}_{xx} && \sigma^{x}_{xz}\\-\sigma^{y}_{xx} && -\sigma^{x}_{xx} && -\sigma^{y}_{xz}\\\sigma^{z}_{xx} && -\sigma^{z}_{xy} && 0 \end{pmatrix}\end{math} & \begin{math} \begin{pmatrix} -\sigma^{y}_{xx} && -\sigma^{x}_{xx} && \sigma^{y}_{xz}\\-\sigma^{x}_{xx} && \sigma^{y}_{xx} && \sigma^{x}_{xz}\\\sigma^{z}_{xy} && \sigma^{z}_{xx} && 0 \end{pmatrix}\end{math} & \begin{math} \begin{pmatrix} \sigma^{x}_{zx} && \sigma^{y}_{zx} && 0\\-\sigma^{y}_{zx} && \sigma^{x}_{zx} && 0\\0 && 0 && \sigma^{z}_{zz} \end{pmatrix}\end{math} & \begin{math} \begin{pmatrix} \sigma^{x}_{xx} && \sigma^{y}_{xx} && \sigma^{x}_{xz}\\\sigma^{y}_{xx} && -\sigma^{x}_{xx} && -\sigma^{y}_{xz}\\\sigma^{x}_{zx} && -\sigma^{y}_{zx} && 0 \end{pmatrix}\end{math} & \begin{math} \begin{pmatrix} \sigma^{y}_{xx} && -\sigma^{x}_{xx} && \sigma^{y}_{xz}\\-\sigma^{x}_{xx} && -\sigma^{y}_{xx} && \sigma^{x}_{xz}\\\sigma^{y}_{zx} && \sigma^{x}_{zx} && 0 \end{pmatrix}\end{math} & \begin{math} \begin{pmatrix} \sigma^{z}_{xx} && \sigma^{z}_{xy} && 0\\-\sigma^{z}_{xy} && \sigma^{z}_{xx} && 0\\0 && 0 && \sigma^{z}_{zz} \end{pmatrix}\end{math}\\[4mm]
 \spacegroup{6'/m'} & \begin{math} \begin{pmatrix} \sigma^{x}_{xx} && \sigma^{y}_{xx} && -\sigma^{x}_{xz}\\\sigma^{y}_{xx} && -\sigma^{x}_{xx} && -\sigma^{y}_{xz}\\-\sigma^{z}_{xx} && -\sigma^{z}_{xy} && 0 \end{pmatrix}\end{math} & \begin{math} \begin{pmatrix} \sigma^{y}_{xx} && -\sigma^{x}_{xx} && \sigma^{y}_{xz}\\-\sigma^{x}_{xx} && -\sigma^{y}_{xx} && -\sigma^{x}_{xz}\\\sigma^{z}_{xy} && -\sigma^{z}_{xx} && 0 \end{pmatrix}\end{math} & \begin{math} \begin{pmatrix} -\sigma^{x}_{zx} && \sigma^{y}_{zx} && 0\\-\sigma^{y}_{zx} && -\sigma^{x}_{zx} && 0\\0 && 0 && -\sigma^{z}_{zz} \end{pmatrix}\end{math} & \begin{math} \begin{pmatrix} \sigma^{x}_{xx} && \sigma^{y}_{xx} && \sigma^{x}_{xz}\\\sigma^{y}_{xx} && -\sigma^{x}_{xx} && -\sigma^{y}_{xz}\\\sigma^{x}_{zx} && -\sigma^{y}_{zx} && 0 \end{pmatrix}\end{math} & \begin{math} \begin{pmatrix} \sigma^{y}_{xx} && -\sigma^{x}_{xx} && \sigma^{y}_{xz}\\-\sigma^{x}_{xx} && -\sigma^{y}_{xx} && \sigma^{x}_{xz}\\\sigma^{y}_{zx} && \sigma^{x}_{zx} && 0 \end{pmatrix}\end{math} & \begin{math} \begin{pmatrix} \sigma^{z}_{xx} && \sigma^{z}_{xy} && 0\\-\sigma^{z}_{xy} && \sigma^{z}_{xx} && 0\\0 && 0 && \sigma^{z}_{zz} \end{pmatrix}\end{math}\\[4mm]
 \spacegroup{6'/m'm'm} & \begin{math} \begin{pmatrix} \sigma^{x}_{xx} && 0 && 0\\0 && -\sigma^{x}_{xx} && -\sigma^{y}_{xz}\\0 && -\sigma^{z}_{xy} && 0 \end{pmatrix}\end{math} & \begin{math} \begin{pmatrix} 0 && -\sigma^{x}_{xx} && \sigma^{y}_{xz}\\-\sigma^{x}_{xx} && 0 && 0\\\sigma^{z}_{xy} && 0 && 0 \end{pmatrix}\end{math} & \begin{math} \begin{pmatrix} 0 && \sigma^{y}_{zx} && 0\\-\sigma^{y}_{zx} && 0 && 0\\0 && 0 && 0 \end{pmatrix}\end{math} & \begin{math} \begin{pmatrix} \sigma^{x}_{xx} && 0 && 0\\0 && -\sigma^{x}_{xx} && -\sigma^{y}_{xz}\\0 && -\sigma^{y}_{zx} && 0 \end{pmatrix}\end{math} & \begin{math} \begin{pmatrix} 0 && -\sigma^{x}_{xx} && \sigma^{y}_{xz}\\-\sigma^{x}_{xx} && 0 && 0\\\sigma^{y}_{zx} && 0 && 0 \end{pmatrix}\end{math} & \begin{math} \begin{pmatrix} 0 && \sigma^{z}_{xy} && 0\\-\sigma^{z}_{xy} && 0 && 0\\0 && 0 && 0 \end{pmatrix}\end{math}\\[4mm]
 \spacegroup{6'/m'mm'} & \begin{math} \begin{pmatrix} 0 && \sigma^{y}_{xx} && 0\\\sigma^{y}_{xx} && 0 && -\sigma^{y}_{xz}\\0 && -\sigma^{z}_{xy} && 0 \end{pmatrix}\end{math} & \begin{math} \begin{pmatrix} \sigma^{y}_{xx} && 0 && \sigma^{y}_{xz}\\0 && -\sigma^{y}_{xx} && 0\\\sigma^{z}_{xy} && 0 && 0 \end{pmatrix}\end{math} & \begin{math} \begin{pmatrix} 0 && \sigma^{y}_{zx} && 0\\-\sigma^{y}_{zx} && 0 && 0\\0 && 0 && 0 \end{pmatrix}\end{math} & \begin{math} \begin{pmatrix} 0 && \sigma^{y}_{xx} && 0\\\sigma^{y}_{xx} && 0 && -\sigma^{y}_{xz}\\0 && -\sigma^{y}_{zx} && 0 \end{pmatrix}\end{math} & \begin{math} \begin{pmatrix} \sigma^{y}_{xx} && 0 && \sigma^{y}_{xz}\\0 && -\sigma^{y}_{xx} && 0\\\sigma^{y}_{zx} && 0 && 0 \end{pmatrix}\end{math} & \begin{math} \begin{pmatrix} 0 && \sigma^{z}_{xy} && 0\\-\sigma^{z}_{xy} && 0 && 0\\0 && 0 && 0 \end{pmatrix}\end{math}\\[4mm]
 \spacegroup{m\bar 3m'} & \begin{math} \begin{pmatrix} 0 && 0 && 0\\0 && 0 && \sigma^{z}_{xy}\\0 && \sigma^{y}_{xz} && 0 \end{pmatrix}\end{math} & \begin{math} \begin{pmatrix} 0 && 0 && \sigma^{y}_{xz}\\0 && 0 && 0\\\sigma^{z}_{xy} && 0 && 0 \end{pmatrix}\end{math} & \begin{math} \begin{pmatrix} 0 && \sigma^{z}_{xy} && 0\\\sigma^{y}_{xz} && 0 && 0\\0 && 0 && 0 \end{pmatrix}\end{math} & \begin{math} \begin{pmatrix} 0 && 0 && 0\\0 && 0 && \sigma^{z}_{xy}\\0 && \sigma^{y}_{xz} && 0 \end{pmatrix}\end{math} & \begin{math} \begin{pmatrix} 0 && 0 && \sigma^{y}_{xz}\\0 && 0 && 0\\\sigma^{z}_{xy} && 0 && 0 \end{pmatrix}\end{math} & \begin{math} \begin{pmatrix} 0 && \sigma^{z}_{xy} && 0\\\sigma^{y}_{xz} && 0 && 0\\0 && 0 && 0 \end{pmatrix}\end{math}\\[4mm]
 \bottomrule
\end{tabular}